\documentclass[12pt]{article}
\usepackage{amsfonts,amsmath,amssymb,epsf}
\usepackage{graphics} 
\usepackage{ifpdf}
\ifpdf
\usepackage{epstopdf}
\usepackage{hyperref}
\else
\usepackage[hypertex]{hyperref}
\fi
\usepackage{xypic}

\advance\textheight by \topskip
\newcounter{defthm}
 
\newtheorem{defthm}{\whattheorem}[section]

\newcommand\sect[1]{\section{#1}\setcounter{equation}0\setcounter{defthm}0}

\newcommand\void[1]       {}

\newcommand\be            {\begin{equation}}
\newcommand\bea           {\begin{equation}\begin{array}l\displaystyle}
\newcommand\bearll        {\begin{array}{ll}\displaystyle}
\newcommand\ee            {\end{equation}}
\newcommand\eear          {\end{array}}
\newcommand\enl           {\\[1em]\displaystyle}
\newcommand\etb           {& \displaystyle}
\newcommand\erf[1] {(\ref{#1})}
\newcommand\labl[1]       {\label{#1}\ee}

\newcommand\bb            {{\underline b}{}}

\renewcommand\cir         {\,{\circ}\,}

\newcommand\eps           {\varepsilon}
\newcommand\Hom           {\text{Hom}}
\newcommand\id            {{\rm id}}
\newcommand\im            {{\rm im}}
\newcommand\In            {\,{\in}\,}
\newcommand\Irr           {\text{Irr}}

\newcommand\oti           {\,{\otimes}\,}
\newcommand\ti            {\,{\times}\,}
\newcommand\VxV           {\Vc{\times}\bar\Vc}

\newcommand\Cb            {\mathbb{C}}

\newcommand\Rb            {\mathbb{R}}
\newcommand\Zb            {\mathbb{Z}}

\newcommand\Cc            {\mathcal{C}}
\newcommand\Nc            {\mathcal{N}}
\newcommand\Vc            {\mathcal{V}}
\newcommand\Wc            {\mathcal{W}}

\newcommand\ev            {{\rm ev}}

\newcommand{\Hc} {\mathcal{H}}

\begin{document}
\thispagestyle{empty}
\def\thefootnote{\fnsymbol{footnote}}
\begin{flushright}
KCL-MTH-07-08\\
0707.0388 [hep-th]
\end{flushright}
\vskip 2.0em
\begin{center}\LARGE
{}From boundary to bulk in logarithmic CFT
\end{center}\vskip 1.5em
\begin{center}\large
  Matthias R. Gaberdiel%
  $^{a}$\footnote{Email: {\tt gaberdiel@itp.phys.ethz.ch}}
  and
  Ingo Runkel%
  $^{b}$\footnote{Email: {\tt ingo.runkel@kcl.ac.uk}}%
\end{center}
\begin{center}\it$^a$
Institut f\"ur Theoretische Physik, ETH Z\"urich \\
8093 Z\"urich, Switzerland
\end{center}
\begin{center}\it$^b$
  Department of Mathematics, King's College London \\
  Strand, London WC2R 2LS, United Kingdom  
\end{center}
\vskip .5em
\begin{center}
  July 2007
\end{center}
\vskip 1em
\begin{abstract}
The analogue of the charge-conjugation modular invariant 
for rational logarithmic conformal field theories is constructed.
This is done by reconstructing the bulk spectrum from a simple
boundary condition (the analogue of the Cardy `identity brane'). 
We apply the general method to the $c_{1,p}$ 
triplet models and reproduce the previously known bulk theory for 
$p=2$ at $c=-2$. 
For general $p$ we verify that
the resulting partition functions are modular invariant. We also
construct the complete set of $2p$ boundary states, and confirm that
the identity brane from which we started indeed exists.  
As a by-product we obtain 
a logarithmic version of the Verlinde formula for
the $c_{1,p}$ triplet models.
\end{abstract}

\setcounter{footnote}{0}
\def\thefootnote{\arabic{footnote}}

\newpage

\sect{Introduction}

In the last few years logarithmic conformal field theories have
increasingly attracted attention. They
appear in various models of statistical
physics, for example in  the theory of (multi)critical polymers
\cite{Sal92,Flohr95,Kau95}, percolation 
\cite{Watts96,Flohr:2005ai,Rasmussen:2007pc},
and various critical (disordered) models 
\cite{CKT95,MSer96,CTT98,Gurarie:1999yx,Ruelle,deGier:2003dg,Piroux:2004vd,%
Gurarie:2004ce,Moghimi-Araghi:2004wg,Jeng:2006tg}. During the last
year also lattice realisations of logarithmic conformal field
theories have been found \cite{Pearce:2006sz,Read:2007qq,RS1}. In a
separate development WZW models on supergroups (that also exhibit
logarithmic behaviour) have been more intensively studied
\cite{RSal92,Schomerus:2005bf,Gotz:2006qp,Saleur:2006tf,Quella:2007hr}. 
These supergroup theories are likely to play an important role for the
world-sheet description of string theory on $AdS$ spaces. Finally,
logarithmic conformal field theories are also interesting from an
abstract point of view since they fall outside the well-studied class
of rational conformal field theories and thus represent a first step
towards understanding at least some aspects of non-rational
theories. The  abstract structure of logarithmic conformal 
field theories has also been
studied, starting from \cite{Gurarie:1993xq,GabKau96a}, and more
recently in the mathematical literature 
\cite{Milas:2001bb,Miyamoto,Fuchs:2003yu,Huang:2003za,%
Feigin:2005xs,Fuchs:2006nx,Huang:2006yz}.
Reviews about different aspects of logarithmic conformal field
theories are \cite{Flohr:2001zs,Gaberdiel:2001tr,Kawai:2002fu}.

While there has recently been some interesting progress with the
supergroup theories, the best understood logarithmic conformal field
theory continues to be the rational triplet theory at $c=-2$
\cite{Kausch91}. It is the only logarithmic 
conformal field theory for which all structures have been understood
in detail from a conformal field theory point of view. In particular,
the fusion rules of this theory were derived from first principles in
\cite{GKau96b}, and a consistent local theory, whose amplitudes
satisfy crossing symmetry, has been constructed by solving the
conformal bootstrap in \cite{Gaberdiel:1998ps}. More recently, a
careful analysis of the boundary theory has been performed in
\cite{Gaberdiel:2006pp}, and a consistent set of boundary states has
been found. (Some of these results were anticipated in
\cite{Kawai:2001ur}, see also
\cite{Kogan:2000fa,Ishimoto:2001jv,Bredthauer:2002ct,Bredthauer:2002xb}
for earlier discussions.) 

The local bulk theory of the triplet theory \cite{Gaberdiel:1998ps} is
actually quite complicated --- its only simple description is in terms
of the symplectic fermions \cite{Kau95,Kausch:2000fu} that are quite
special for the $c=-2$ triplet theory --- but the boundary theory of  
\cite{Gaberdiel:2006pp} turned out to be remarkably simple. The
boundary states are labelled by the irreducible representations of the
triplet theory, while the open string spectrum consists precisely of
the representations that appear in the fusion of these 
irreducible representations. (These representations involve then
in general indecomposable logarithmic representations.) In this paper
we show how the bulk theory can actually be obtained in a very natural
manner from the boundary theory. Our analysis applies to all
rational logarithmic conformal field theories; for the case of the
$c_{1,p}$ models we can describe the resulting bulk theory very
explicitly, and for $p=2$ it coincides with the original $c=-2$ bulk
theory of \cite{Gaberdiel:1998ps}. 
\medskip

Our method is based on insights into the general structure
of conventional (non-lo\-ga\-rith\-mic) rational conformal field
theories that have been obtained during the last few years 
\cite{Fuchs:2002cm,Fuchs:2004xi,Fjelstad:2005ua,Fjelstad:2006aw}. 
In particular, it
has become clear that a good way to describe a given rational
conformal field theory is by starting from its boundary theory: given
the spectrum of boundary fields on a single boundary condition (that
preserves the full chiral algebra) as well as the associative 
operator product of these fields, one can reconstruct the bulk theory
for which this boundary theory describes an
allowed boundary condition. In particular, this allows
one to solve the complicated conformal bootstrap equations in terms of
the much simpler problem of constructing an associative operator
product on the boundary.

The basic idea behind this reconstruction can be described as
follows. One can argue on general grounds that the disc correlation 
functions that involve one bulk field and one boundary field 
are non-degenerate in the bulk field insertion. This
is to say, for any non-trivial bulk field there exists a boundary
field such that the corresponding disc correlator does not
vanish. Knowing the boundary spectrum thus gives constraints on the
possible size of the bulk theory. Furthermore, the correlation
functions involving one bulk field and
two boundary fields must essentially be
independent of the order in which the latter 
appear on the boundary since
one can take one of them around the 
circle (see figure~2 below). It was shown in
\cite{Fuchs:2002cm,Fjelstad:2006aw} that in the non-logarithmic
rational case the bulk theory is then simply the largest 
possible representation of the two chiral algebras that satisfies
these two constraints. Furthermore one can determine from this data
also the bulk correlation functions, {\it etc}.

While the corresponding statement is not yet known for the logarithmic
case (that falls outside the mathematical analysis of
\cite{Fuchs:2002cm,Fjelstad:2006aw}), it is clear that these two 
conditions also have to hold in the general (logarithmic)
situation. We can therefore use these ideas to constrain the possible
spectrum of the bulk theory starting from a boundary condition. Given
the above observations about the boundary theory of the $c=-2$ triplet
model, it seems very likely that at least all $c_{1,p}$ triplet models
will have a boundary condition whose boundary spectrum consists just
of the vacuum representation of the triplet algebra. (This is the
boundary condition associated to the irreducible vacuum
representation.) Starting from such a boundary condition we can 
analyse the above constraints and
construct the largest space that is compatible with them. For the
specific case of the $c_{1,p}$ triplet models (for which some aspects
of the allowed representations are known) we can then be even more
specific and describe the resulting bulk space very explicitly,
see \erf{eq:Wp-bulk-space}. 
As we show in detail, it leads to a modular invariant partition
function and supports boundary conditions in one-to-one correspondence
with the irreducible representations of the triplet algebra. This
gives strong support to the assertion that this describes in fact the
correct bulk theory. It also reduces to the known bulk theory
\cite{Gaberdiel:1998ps} for $c=-2$ and is compatible with
the predictions (based on the analogy with supergroups) of
\cite{Quella:2007hr}. 

In the usual rational case, the bulk theory corresponding to the
boundary condition whose spectrum consists just of the vacuum
representation itself, is the charge-conjugation modular
invariant \cite{Runkel:1998pm,Felder:1999mq}. The above theories
should therefore be thought of as the analogue of the
charge-conjugation construction. One may suspect that there will also
be other consistent bulk theories (with other modular invariant
partition functions \cite{Flohr:2001zs,Fuchs:2003yu,Feigin:2006iv}). 
It would be interesting to study this for the
example of the triplet theories. 

As a by-product of our analysis we find an expression for the
boundary states of the $c_{1,p}$ models in terms of the 
$S$-matrices. Since the open string multiplicities are determined in
terms of the fusion rules, this then leads to a Verlinde-like formula
for the fusion rules of these models. (More precisely, the formula
describes the product in the associated Grothendieck ring.) 
Given the abstract form of the formula it is very natural 
to suspect that it will generalise to other logarithmic 
rational conformal field theories.
\medskip

The paper is organised as follows. In section~\ref{sec:constr-bulk} 
we explain the general
method of reconstructing the bulk theory from a given boundary
condition. In section~\ref{sec:identity-brane} 
we concentrate on the case that the boundary
only has the vacuum representation in its boundary spectrum
and derive the constraints on the possible bulk space in the general
logarithmic case. We also give a fairly explicit description of the
largest such space. In section~\ref{sec:Wp-minmod}
these ideas are then applied to the
$c_{1,p}$ triplet models. In particular, we give a detailed
description of the bulk spectrum for general $p$ in 
section~\ref{sec:bulk-construct}, and
show that it reproduces the known result for $p=2$. We also show 
in section~\ref{sec:Wp-mod-inv} that
it leads to a modular invariant partition function. 
Finally, in section~\ref{sec:boundary-states}
we analyse the Cardy condition for
this bulk spectrum, and show how to construct boundary states in
one-to-one correspondence with the irreducible
representations. We also discuss the Verlinde formula for logarithmic
rational conformal field theories there. 
Section~\ref{sec:concl} 
contains our conclusions. There are a
number of appendices in which some of the more technical material is
described. Among other things, we also conjecture there
the fusion rules for the general 
${\mathcal W}_p$ triplet algebras at $c=c_{1,p}$,
see \erf{eq:irrep-fusion}, \erf{eq:fusion-expl} and \erf{eq:fusion2}.

\sect{Constructing the space of bulk states}\label{sec:constr-bulk}

In this section we will generalise one key element of 
non-logarithmic rational conformal field theories to the
logarithmic case, namely the construction of the space of bulk fields
from a given algebra of boundary fields \cite{Fuchs:2002cm}. 

\subsection{The bulk space from disc amplitudes}

Suppose we are given a conformal field theory (logarithmic or not)
that is defined on surfaces with (and without) boundaries. In
particular the theory is defined on the unit disc, where at the 
boundary of the disc we have chosen one of the possible boundary
conditions (that we shall denote by $\gamma$). Consider now the
correlator involving an arbitrary bulk field in the interior of the
disc, together with a single boundary field on the boundary. By the
usual ${\rm SU}(1,1)$  symmetry of disc correlation functions, we may
assume without loss of generality that the bulk field is inserted at
$z=0$, while the boundary field sits at $z=1$. This correlator defines
a bilinear pairing 
\be
b : \Hc_\text{bulk} \times \Hc_\text{bnd} \rightarrow \Cb \ , \qquad 
b(\phi,\psi) = \langle \phi(0) \psi(1) \rangle_\gamma\ ,
\ee
where $\Hc_\text{bulk}$ is the space of bulk fields, while
$\Hc_\text{bnd}$ denotes the space of boundary fields on the boundary
$\gamma$.

\begin{figure}[bt]
\begin{center}
  \begin{picture}(400,100)
    \put(0,0){ \scalebox{.8}{\includegraphics{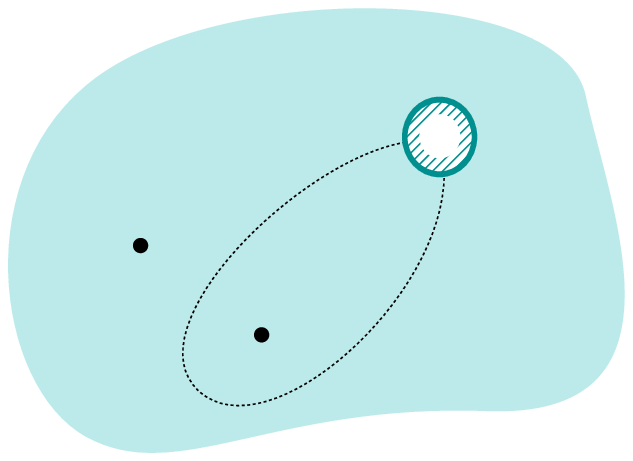}} }
    \put(205,17){ \scalebox{.8}{\includegraphics{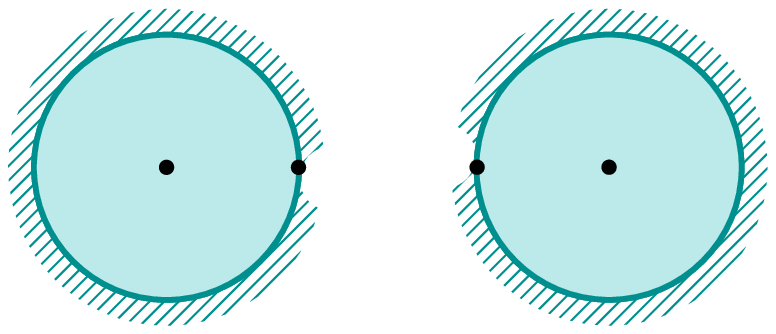}} }
    \put(163,50){$\displaystyle = ~\sum_\alpha$}
    \put(0,0){
     \setlength{\unitlength}{.8pt}\put(-16,-8){
     \put( 84, 50)  {\scriptsize$ \phi'(w) $}
     \put( 50, 76)  {\scriptsize$ \phi(z) $}
     \put(143, 98)  {\scriptsize$ p $}
     }\setlength{\unitlength}{1pt}}
    \put(205,17){
     \setlength{\unitlength}{.8pt}\put(-8,-14){
     \put(178, 68)  {\scriptsize$ \phi'(0) $}
     \put( 50, 68)  {\scriptsize$ \phi(0) $}
     \put(102, 55)  {\scriptsize$ \psi_\alpha $}
     \put(132, 61)  {\scriptsize$ \bar\psi_\alpha $}
     }\setlength{\unitlength}{1pt}}
  \end{picture}
\end{center}
\caption{The correlator of two bulk fields on the complex plane with a
  little hole can be written as a sum of products of disk correlators
  by factorising along the dashed interval. (All factors arising from
  the conformal transformations to the disc have been absorbed into
  the bases $\psi_\alpha$ and $\bar \psi_\alpha$.)} 
\label{fig:bulk-disk}
\end{figure}

We will now argue that this pairing is non-degenerate in the first
argument. This means that for any non-zero bulk field $\phi$, there
exists a boundary field $\psi$ such that the correlator does not
vanish, $b(\phi,\psi)\neq 0$. To see this we recall that, by
definition, the two-point functions on the sphere define a
non-degenerate bilinear pairing on the space of bulk fields. (This is
to say, if a bulk field $\phi$ vanishes in all two-point functions on
the sphere, then we have in fact $\phi=0$.) This property should not
change if we consider instead the two-point
function on the sphere with a little boundary circle
around some point $p$ far away from the insertion points of the bulk
fields. But then we can use factorisation along an interval starting
and ending on this boundary circle to express the correlation function
as a sum over products of disc correlators, see
figure \ref{fig:bulk-disk}. It is then clear that the bulk-boundary
correlators must be non-degenerate in the bulk fields in order for the
above two-point function
to be non-degenerate. This proves that the bilinear
pairing $b$ is non-degenerate with respect to the first argument. We
note in passing that the argument does not imply that $b$ must be
non-degenerate in the boundary fields as well; in fact, this is not
true in general. (Consider for example a superposition of boundary
conditions and take $\psi$ to be a boundary changing operator. Then
$b(\phi,\psi)=0$ for all bulk fields $\phi$.) 

If we are given a boundary condition $\gamma$ with some space of
boundary fields $\Hc_\text{bnd}$, the condition that the bulk-boundary
correlation functions must be non-degenerate in the bulk fields will
give restrictive constraints on the structure of the bulk space
$\Hc_\text{bulk}$. This will in particular be the case if
$\Hc_\text{bnd}$ is rather small, for example if it just consists of
the chiral algebra $\Vc$ of the theory itself. One can then turn the
logic around and `reconstruct' the bulk space from the boundary
condition. This is what we shall be doing in the following. First,
however, we briefly want to elaborate on the general situation.

\subsection{Constraints on the bulk space from a generic brane}

We denote the chiral algebra of the bulk theory ({\it i.e.}\ the
conformal vertex algebra of the holomorphic degrees of freedom) 
by $\Vc$, and by $\VxV$ the holomorphic and anti-holomorphic copy of
$\Vc$ in the bulk. We shall always consider boundary conditions
that preserve $\Vc$; thus we assume that for any holomorphic field $W$
of $\Vc$ we have 
\be\label{gluing}
W(z) = \bar{W} (\bar{z}) \ , \qquad z=\bar{z} \ ,
\ee
where $\bar{W}$ is the corresponding field in $\bar\Vc$. (We have
written this condition for the case where instead of the disk we are
considering the upper half plane with boundary the real axis.) 
The space of boundary fields $\Hc_\text{bnd}$ is then a representation
of $\Vc$; as usual, the operator product on $\Hc_\text{bnd}$ must 
be associative. 

\begin{figure}[bt]
\begin{center}
  \begin{picture}(280,100)
    \put(0,10){ \scalebox{1}{\includegraphics{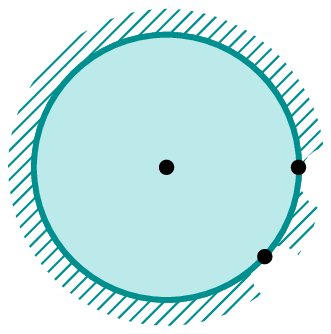}} }
    \put(160,0){ \scalebox{1}{\includegraphics{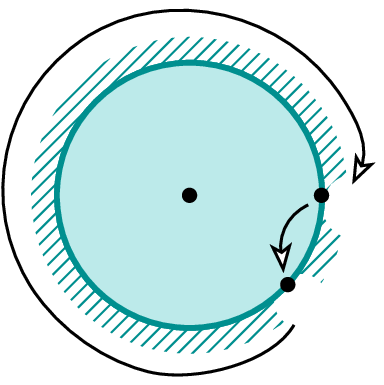}} }
    \put(130,50){$ = $}
    \put(0,10){
     \setlength{\unitlength}{1pt}\put(-8,-14){
     \put( 63, 55)  {\scriptsize$ \phi(0) $}
     \put(101, 55)  {\scriptsize$ \psi(\theta_1) $}
     \put( 88, 26)  {\scriptsize$ \psi'(\theta_2) $}
     }\setlength{\unitlength}{1pt}}
    \put(160,0){
     \setlength{\unitlength}{1pt}\put(0,-7){
     \put( 63, 55)  {\scriptsize$ \phi(0) $}
     \put(101, 55)  {\scriptsize$ \psi'(\theta_1) $}
     \put( 88, 26)  {\scriptsize$ \psi(\theta_2) $}
     }\setlength{\unitlength}{1pt}}
  \end{picture}
\end{center}
\caption{A constraint on the possible bulk fields $\phi$: Inserting two
  boundary fields in reversed order is equivalent to analytic
  continuation around the disk. The two disk correlators are fixed
  separately by $b$ and the operator product expansion on
$\Hc_\text{bnd}$.}  
\label{fig:cont-around-disk}
\end{figure}

The arguments of the previous subsection now imply that the space of
bulk fields $\Hc_\text{bulk}$ must have the property that:  
\begin{list}{-}{\topsep .2em \leftmargin 2em}
\item[(1)]
There exists a pairing 
$b : \Hc_\text{bulk} \times \Hc_\text{bnd} \rightarrow \Cb$ compatible
with the action of $\Vc$ and non-degenerate in the first argument.  
\end{list}
The compatibility condition with the $\Vc$-action follows from the
usual contour deformation arguments involving the holomorphic fields
in $\Vc$; it will be given in more detail in section \ref{sec:universal} 
and appendix~\ref{app:b-mode-com}. 
Using the associativity of the operator product expansion on
the boundary, the pairing $b$ then also determines uniquely the disk
correlator of an arbitrary number of boundary fields with one bulk
field. A second constraint is then:  
\begin{list}{-}{\topsep .2em \leftmargin 2em}
\item[(2)]
A disk correlator with bulk insertion $\phi(0)$ 
and boundary insertions $\psi(\theta_1)\psi'(\theta_2)$ 
has to be related to the correlator with reversed 
boundary insertions $\psi'(\theta_1) \psi(\theta_2)$ by analytic
continuation, see figure \ref{fig:cont-around-disk}.  
\end{list}
This second condition is just one of the sewing constraints for
conformal field theories with boundary
\cite[fig.\,9d]{Lewellen:1991tb}. For rational conformal field
theories, using the language of \cite[sect.\,5.3]{Fuchs:2002cm}, it
amounts to the statement that bulk fields are in the image of a
certain projector, while in the approach of \cite{Kong2006} it is
definition 5.11.  

For non-logarithmic rational conformal field theories 
one can show that $\Hc_\text{bulk}$ is uniquely 
determined by the boundary condition $\gamma$ ({\it i.e.}\ by the 
associative algebra of boundary fields on $\gamma$), and that it is
simply the largest $\VxV$-representation that satisfies these 
constraints (see \cite[lemma~5.6]{Fuchs:2002cm} and \cite{Fjelstad:2006aw}). 
While we do not yet know how to prove the
corresponding statement in the general logarithmic case, it is clear
that any consistent $\Hc_\text{bulk}$ must satisfy at least these
constraints. Furthermore the examples we shall study  
below suggest that
$\Hc_\text{bulk}$ is again (also in the logarithmic case) simply the 
largest $\VxV$-representation that satisfies (1) and (2).

\section{The identity brane}\label{sec:identity-brane}
\setcounter{equation}{0}

We now want to discuss the construction of the bulk space for the
simplest case where the boundary spectrum consists just of the vacuum
representation of the chiral algebra, {\it i.e.}\ for which 
$\Hc_\text{bnd}=\Vc$. In the non-logarithmic 
rational case such a brane exists in the charge conjugation theory,
namely as the Cardy brane associated to the vacuum representation
\cite{Cardy:1989ir}. For
the logarithmic triplet model at $c=-2$ for which the boundary
conditions were analysed in detail in \cite{Gaberdiel:2006pp}, we also
found one such brane. 

In the following we shall thus assume that we have a boundary
condition $\gamma$ for which $\Hc_\text{bnd}=\Vc$. 
We want to construct a bulk theory $\Hc_{\text{bulk}}$ that satisfies
conditions (1) and (2) relative to this boundary. Since 
$\Hc_\text{bnd}=\Vc$, condition (2) is 
simply implied by the fact that $\Hc_\text{bulk}$ is 
a representation of $\VxV$. Thus we only need to find a solution to 
condition (1). 
To this end we start with some large space of potential 
bulk states $\hat\Hc$. We then calculate the correlation functions of
bulk states in $\hat\Hc$ on the disc with the boundary condition
$\gamma$; these are determined by the chiral symmetry up to some
coupling constants (normalisations of three-point blocks). For any
choice of these coupling constants we then find the subspace 
$\Nc\subset \hat\Hc$ of potential bulk states that vanishes in
all such disc correlation functions; for the given choice of coupling
constants the actual bulk space is thus the quotient 
$\Hc_{\text{bulk}} = \hat\Hc / \Nc$. Obviously, the null-space 
$\Nc$ (and therefore $\Hc_\text{bulk}$) depends on the choice of
these coupling constants, but as we shall see, the resulting space is 
essentially independent of these choices as long as we pick generic
values. We thus define the bulk space to be the largest such
space as we vary the coupling constants. 
For the $c_{1,p}$ triplet models we will see in 
sections \ref{sec:Wp-minmod} and \ref{sec:boundary-states} 
that the resulting space leads to a modular
invariant partition function and gives rise to the expected boundary
states, in particular one with $\Hc_\text{bnd}=\Vc$. The fact that 
this last boundary condition satisfies the Cardy constraint is not a
priori guaranteed, and hence provides a consistency check on
our approach.

After this informal description of the strategy, we now want to
give more details of the construction.

\subsection{The universal property defining $\Hc_\text{bulk}$}
\label{sec:universal}

As we have just mentioned we shall from now on assume that 
$\Hc_{\text{bnd}}$ is just the chiral algebra itself
$\Hc_\text{bnd}=\Vc$. 
Let us start with a simple ansatz for the space
of potential bulk states $\hat\Hc$, namely that $\hat\Hc$ is  
the direct sum of tensor products of representations of $\Vc$
and $\bar\Vc$; a particular term in this sum will thus be of the form
$M\otimes \bar{N}$, where $M$ and $N$ are representations of
$\Vc$. (The bar on $N$ indicates that it describes the right-moving
degrees of freedom that form a representation of $\bar\Vc$.)
A disk correlator with one bulk insertion in $M\otimes \bar{N}$
at $z=0$ and a boundary 
insertion at $z=1$ can be mapped conformally to the upper half plane
with a boundary insertion at $0$ and a bulk insertion at $i$. Since
the boundary condition preserves the chiral algebra (\ref{gluing}) we
can use the doubling trick \cite{Cardy:1984bb} to write this
correlator as the three-point block on the complex plane with
an insertion of $\Vc$ at $0$, while $M$ and $N$ are inserted at 
$\pm i$. Every such conformal block $\beta$ defines a multilinear map
$M \times N \times \Vc \rightarrow \Cb$ that obeys  
invariance conditions with respect to the $\Vc$-actions which are
listed explicitly in appendix~\ref{app:b-mode-com}. Furthermore, 
the three-point block $\beta$ gives rise to a bulk-boundary correlator,
and thus to an associated pairing 
$b_\beta: (M\oti\bar N)\times \Vc \rightarrow \Cb$, 
$(a \oti b,v) \mapsto \beta(a,b,v)$.

Similarly, if $\hat\Hc = \bigoplus_k M_k \oti \bar N_k$ then the pairings 
that are compatible with the $\Vc$-action are of the form 
$\sum_k b_{\beta_k}$, where $\beta_k$ is a three-point block  
$M_k \times N_k \times \Vc \rightarrow \Cb$. In fact, the possible
pairings can also be described for an arbitrary $\VxV$-representation
$\hat\Hc$, not just one of the form $\bigoplus_{k} M_k \oti \bar N_k$;
as is shown in  appendix~\ref{app:b-mode-com}, 
they have to satisfy condition \erf{eq:bilin-Uhat-V-cond}.
We denote the space of all such pairings by
$B(\hat\Hc)$.    

Since we are only interested in the non-degeneracy of $b$ in the first
(bulk) entry, it is convenient to associate to each 
$b \in B(\hat\Hc)$ the map $\bb: \hat\Hc \rightarrow \Vc^*$ given by  
$\bb(\phi) = {b}(\phi,\,{\cdot}\,)$. The condition (1) of the previous
section is then just the requirement that $\bb$ must be injective
({\it i.e.}\ that its kernel is trivial). In general, this will not be
the case for our ansatz $\hat\Hc$, but it is easy to rectify this
problem. We denote the kernel of $\bb$ by 
$\Nc = \ker(\bb) \subset \hat\Hc$. Then for the quotient space  
$\Hc = \hat\Hc/\Nc$ the induced pairing on
$\Hc \times \Vc$ is by 
construction non-degenerate in the first argument. As is shown in
appendix \ref{app:kernel-of-b}, $\Hc$ is still a representation of 
$\VxV$ since $\Nc$ is; this 
is for example necessary to guarantee that
condition (2) continues to hold.

Obviously, the space $\Hc$ we end up with depends to a certain extent
on the choice of the three-point blocks $\beta_k$. (For example, we
could take all $\beta_k=0$, in which case $\Hc$ would be the zero
space.) We expect that the actual bulk space that contains the
boundary condition in question is as large as it can be. In order to
make this precise we need to define what we mean by a `maximal
solution'. A $\VxV$ representation $\Hc$ together with a pairing
$b\in B(\Hc)$ is a maximal solution to (1) if and only if 
$\bb$ is injective and the following universal property holds:  
for any pair $(\Hc_1,{b_1})$ 
such that $b_1 \in B(\Hc_1)$ and
$\bb_1 : \Hc_1 \rightarrow \Vc^*$ is injective there exists a
unique injective intertwiner $f : \Hc_1 \rightarrow \Hc$
such that the following diagram commutes. 
\be
\raisebox{-2em}{
\begin{xy} 
(0,20)*+{\Hc}="a"; (25,20)*+{\Vc^*}="b";%
(0,0)*+{\Hc_1}="c";%
{\ar@{->} "a";"b"}?*!/_2mm/{\bb}; 
{\ar@{-->} "c";"a"}?*!/_4mm/{\exists! f};%
{\ar@{->} "c";"b"};?*!/_2mm/{\bb_1~}; 
\end{xy}}
\labl{eq:maximal-universal}
In fact, if an intertwiner $f$
exists, it is  automatically unique and injective because both $\bb$
and $\bb_1$ are injective. Furthermore, a maximal solution to
condition (1) is unique up to isomorphism.

\subsection{The bulk space in terms of projective 
covers}\label{sec:bulk-proj} 

What we have said so far is completely general, but in order to be
more specific we need to assume some properties about the
representations of the chiral algebra $\Vc$ (more precisely, the
logarithmic modules \cite{Milas:2001bb}, or generalised modules 
\cite[sect.\,2]{Huang:2006yz}). We assume that 
\begin{list}{-}{\topsep .2em \leftmargin 3em \itemsep -.1em}
\item[(i)] $\Vc$ has only finitely many inequivalent irreducible
representations. 
\item[(ii)] Each $\Vc$-representation $M$ has a projective cover  
$P(M)$. 
\end{list}
As will become clear in section~\ref{sec:Wp-minmod}, the  triplet algebras 
${\cal W}_p$ for $p\geq 2$ satisfy these conditions. More abstractly,
one may expect that the representation category of a rational
logarithmic conformal field theory is described by a 
finite tensor category, see {\it e.g.}\
\cite{Etingof2003,Fuchs:2006nx}; then these conditions are
automatically satisfied.
We shall also use the following properties of 
$\VxV$-representations\footnote{
These should hold for reasonable chiral algebras $\Vc$. For example 
(II) follows if the intertwiner spaces $\Hom_{\Vc}(M,N)$ 
are finite-dimensional, and the generalised $L_0$-eigenspaces 
of the $\Vc$-representations are finite-dimensional.}
\begin{list}{-}{\topsep .2em \leftmargin 3em \itemsep -.1em}
\item[(I)] Every $\VxV$-representation $X$ is isomorphic to a quotient
of the tensor product (over $\Cb$) $M_X \oti \bar N_X$ 
of two $\Vc$-representations $M_X$ and $N_X$ by a subrepresentation.
\item[(II)] The space of intertwiners 
   $\Hom_{\VxV}(M \oti \bar N, M' \oti \bar N')$ is isomorphic to 
   the tensor product of intertwiner spaces 
   $\Hom_\Vc(M,M') \oti \Hom_\Vc(N,N')$.  
\end{list}
\medskip

With these preparations we now proceed as follows. As our starting
point we take the space of potential bulk states to be 
\be\label{hpp}
  \hat \Hc = \bigoplus_{k \in \Irr} P_k \oti \bar P_k^* \ ,
\ee
where $\Irr$ labels the finitely many irreducible representations
$U_i$, $i\in\Irr$ of $\Vc$. Here $P_k$ is the projective cover of
$U_k$, and $P_k^*$ is the conjugate representation to $P_k$. 
(More precisely, $P_k$ is the contragredient module to $P_k$ --- for a 
definition see {\it e.g.}\ \cite[sect.\,2]{Huang:2006yz}.) The bar
over the second space in the tensor product indicates that these
degrees of freedom refer to right-movers.

We also need to make an ansatz for the pairing $b$, or equivalently
for the three-point blocks $\beta_k$. In fact, there is an almost
canonical choice we can make: the space of three-point blocks
involving any $\Vc$-representation $M$, its dual representation
$M^\ast$ and $\Vc$ contains a preferred element that we shall denote
by  $\ev_{\!M}$. To define $\ev_M$ we use a conformal transformation 
to move  the insertion points of the three-point block such that  
$M^*$ is inserted at infinity, while $\Vc$ is inserted at $z=1$ and 
$M$ at $z=0$. The three-point block $\ev_{\!M}$ is then uniquely
determined by the condition that upon inserting the vacuum vector
$\Omega$ at $z=1$, the resulting paring 
$M^* \times M \rightarrow \Cb$ is just the canonical pairing of a
vector space with its dual. (The invariance conditions of appendix~A
then determine $\ev_{\!M}$ for any other combination of 
states.) 

For the case at hand $M=P_k$, and we can thus define the pairing $b$
on $\hat\Hc \times \Vc$ to be given by 
\be
  b_{\ev} = b_\beta \qquad \text{with} \quad
  \beta = \sum_{k \in \Irr} \ev_{\!P_k} \ .
\ee
The kernel $\Nc_{\ev}$ of $\bb_{\ev}$ is non-trivial in general,
but as we shall see, the resulting quotient space  
\be
\Hc_{\text{bulk}} = \hat\Hc / \Nc_{\ev}
\labl{eq:bulk-general}
will define a maximal solution to (1). Before we can prove this
statement, we need to make a few observations about a certain class of
three-point blocks.  

\subsection{Three-point blocks and the kernel of $\bb_{\ev}$}
\label{sec:kernel-bev}

Suppose we have a three-point block
$\beta : M \times N \times \Vc \rightarrow \Cb$. Since one of the
three representations (namely $\Vc$) 
is just the vacuum representation, every
such three-point block defines a linear map
$\beta^\sharp : N \rightarrow M^*$ that intertwines the action of
$\Vc$, {\it i.e.}\ satisfies 
$\beta^\sharp \circ W_n = W_n \circ \beta^\sharp$ for every $W_n$ in
$\Vc$. It is then clear that we can write $\beta$ as 
\be\label{eq:ev-gives-all}
\beta(m,n,v) = \ev_{\!M}\big(m,\beta^\sharp(n),v\big) \ , 
\ee
where $m \in M$, $n \in N$ and $v \in \Vc$. 

Similarly, if we have an intertwiner $g$ of the chiral algebra  
mapping the $\Vc$-representations $M$ to $N$, then the 
three-point blocks $\ev_{\!M}$ and $\ev_{\!N}$ are related as 
\be\label{eq:move-g-in-ev}
\ev_{\!N}\Bigl( g(m), n^* , v\Bigr) = 
\ev_{\!M}\Bigl(m,g^*(n^*),v\Bigr) \ , 
\ee
where $g^* \in \Hom_\Vc(N^*,M^*)$ is the linear map dual to $g$. Here 
$m \In M$, $n^* \In N^*$ and $v\In\Vc$.  Equation
(\ref{eq:move-g-in-ev}) can be verified by taking the insertion points
to $0$, $1$ and  $\infty$, and noting that $g$ commutes in particular
with the modes of the Virasoro algebra. 
\smallskip

We are now in a position to give a good description of the kernel of 
$\bb_\ev$, $\Nc_\ev = \ker(\bb_\ev)$. We want to describe it as the
span of the images of intertwiners
$g:P_k\oti \bar P_l^* \rightarrow \hat\Hc$; this is always possible
since by (I) any $\VxV$-representation can be written as a quotient
space of a suitable tensor product which --- by passing to projective
covers --- we may decompose into a direct sum of tensor products of
indecomposable
projective representations. The image of such a map lies in $\Nc_\ev$
if and only if $\bb_\ev \cir g =0$. Thus we can write  
\be
\Nc_\ev = \ker(\bb_\ev)
  = \text{span}_\Cb \big\{ \, 
  \im(g) \, \big|\, 
  g : P_k \oti \bar P_l^* \rightarrow \hat\Hc
  ~\text{with}~ \bb_\ev \cir g\,{=}\,0\ ,\;\; k,l \In \Irr \, \big\}\ .
\labl{eq:ker-bev-v1}
To characterise the relevant intertwiners $g$, we denote by $H_{kl}$
the vector space   
\be\label{Hkl}
H_{kl} = \bigoplus_{i \in \Irr} \Hom_\Vc(P_k,P_i) \oti
\Hom_\Vc(P_i,P_l) \ , 
\ee
where $k,l \in \Irr$ and $\Hom_\Vc(P_r,P_s)$ is the space of
intertwiners from $P_r$ to $P_s$. On $H_{kl}$ we define two maps as
follows. First we have the composition map 
$c_{kl}: H_{kl} \rightarrow  \Hom_\Vc(P_k,P_l)$ which acts on each
component $f_i \oti g_i$ by composition
\be
c_{kl} (f_i \oti g_i) = g_i \circ f_i \in \Hom_\Vc(P_k,P_l) \ .
\labl{eq:cdef}
In addition we have the map
$d_{kl}:H_{kl} \rightarrow \Hom_{\VxV}(P_k \oti \bar{P}_l^*,\hat\Hc)$
defined by setting  
\be
d_{kl} (f_i \oti g_i) = f_i \oti \bar g_i^\ast \in 
\Hom_{\VxV}(P_k \oti \bar{P}_l^*,\hat\Hc) \ , 
\ee
where $g_i^*\in \Hom_\Vc(P_l^*,P_i^*)$ is the dual map to 
$g_i \in \Hom_\Vc(P_i,P_l)$ (and the bar indicates again that 
$g_i^*$ acts now on the right-movers). Note that $d_{kl}$ is an
isomorphism of vector spaces, as follows from (II) above. The key
observation is now that \erf{eq:move-g-in-ev} implies
\bea\label{eq:zentral1}
\ev_{\!P_l} \Bigl([c_{kl}(f_i \oti g_i)] (w_k), \bar{w}_l , v
\Bigr) \equiv  
\ev_{\!P_l} \Bigl([g_i\circ f_i] (w_k), \bar{w}_l , v
\Bigr) \\
\qquad \qquad \qquad = 
\ev_{\!P_i} \Bigl(f_i(w_k), \bar{g}_i^\ast (\bar{w}_l),v\Bigr)
\equiv 
b_{\ev} \Bigl(\Bigl[d_{kl}(f_i\oti g_i)\Bigr] 
(w_k \otimes \bar{w}_l),v \Bigr) \ , 
\eear\ee
where $w_k\In P_k$, $\bar{w}_l\In \bar{P}_l^\ast$ 
and $v \In \Vc$ are arbitrary. 
It therefore follows that 
\be
\bb_{\ev} \circ d_{kl}(F) = 0 \qquad \hbox{if and only if} \qquad  
F \in \ker(c_{kl}) \ .
\ee
Since $d_{kl}$ is an isomorphism, every map $g$ in
\erf{eq:ker-bev-v1} can be written as $d_{kl}(F)$ for an
appropriate $F \in H_{kl}$. Hence the expression \erf{eq:ker-bev-v1} 
for the kernel of $\bb_{\ev}$ can be rewritten as
\be
\Nc_\ev =
  \text{span}_\Cb \big\{ \, \im\big(d_{kl}(F)\big) \, \big|\, 
  F \in \ker(c_{kl})\ ,\;\; k,l \in \Irr \, \big\}\ .
\labl{eq:b1-kernel-general}
The space of bulk states is then defined as in
(\ref{eq:bulk-general}). We also denote by 
$b_\text{disc}$ the pairing on $\Hc_\text{bulk} \times \Vc$ induced by
$b_{\ev}$. It is shown in section \ref{sec:maximal-proof} below that 
$(\Hc_\text{bulk},b_\text{disc})$ is in fact maximal.   
\smallskip

This completes our construction of the bulk space corresponding to the
identity brane. For non-logarithmic rational conformal field theories
we have $P_i = U_i$ 
and one easily verifies that the linear maps $c_{kl}$ all have
trivial kernel. Thus one recovers the space of bulk states of the
charge-conjugation modular invariant theory, 
$\Hc_\text{bulk} = \bigoplus_{k \in \Irr} U_k \oti \bar U_k^*$. 
We shall show in section~\ref{sec:W2-bulk-space} that for the case of the
$c=-2$ triplet theory, the above construction reproduces the known
bulk spectrum \cite{Gaberdiel:1998ps}. We shall also see that it leads
to a very natural bulk spectrum for the other $c_{1,p}$ triplet models
that is in particular modular invariant. 
\smallskip

Before turning to the proof that $(\Hc_\text{bulk},b_\text{disc})$ is
in fact maximal we want to show that it defines at least a local
theory. 

\subsection{Locality}

Locality of the bulk theory requires that the operator 
$\exp( 2 \pi i(L_0{-}\bar L_0))$ acts as the identity on
$\Hc_\text{bulk}$. We want to show now that this is requirement is
automatically satisfied by the above construction. 

First we note that $e^{2 \pi i L_0}$ commutes with all 
generators of $\Vc$ and that it therefore 
defines an intertwiner from any $\Vc$-representation to
itself. Consider now the element  
\be
  t = e^{2 \pi i L_0} \otimes \id - \id \otimes e^{2 \pi i L_0} 
  \in \Hom_\Vc(P_k,P_k) \oti \Hom_\Vc(P_k,P_k) \subset H_{kk} \ .
\ee
It is obvious that $c_{kk}(t) = 0$ and hence $t$ lies in the kernel of
$c_{kk}$. It then follows from (\ref{eq:b1-kernel-general}) 
that the image of $P_k \oti \bar P_k^*$ under
$e^{2 \pi i L_0} \otimes \bar\id - \id \otimes e^{2 \pi i \bar L_0}$ 
lies in $\Nc_\ev$, for all $k \in \Irr$.
In the quotient space $\Hc_\text{bulk}$ we therefore have 
$e^{2 \pi i L_0} = e^{2 \pi i \bar L_0}$, which yields the desired
answer upon acting on both sides with $e^{-2 \pi i \bar L_0}$.

The above argument implies in particular that the torus partition
function for $\Hc_\text{bulk}$ is invariant under 
$\tau \mapsto \tau{+}1$. We do not have a general proof that it is
also invariant under $\tau \mapsto -1/\tau$, but we shall be able to
show the full modular invariance for the $c_{1,p}$ triplet theories
(see section~\ref{sec:Wp-mod-inv}).

\subsection{Proof of maximality}\label{sec:maximal-proof} 

We will now prove that the pair $(\Hc_\text{bulk},b_\text{disc})$ has
the universal property \erf{eq:maximal-universal}.  
Let $(\Hc_1,b_1)$ 
be any pair such that 
$b_1 \in B(\Hc_1)$ and $\bb_1$ is injective. Because of (I)
above, there exist projective representations $P$ and $Q$, as well as
a subrepresentation $K$ of $P \oti \bar Q$ such that 
$\Hc_1 \cong (P \oti \bar Q)/K$. Let 
$\pi_1 : P \oti \bar Q \rightarrow \Hc_1$ be the corresponding  
projection, and set $\underline d = \bb_1 \cir \pi_1$. Since
$\bb_1$ is injective, we have $\ker(\underline d) = K$.

Next we write $P$ in terms of indecomposable projectives as
$P = \bigoplus_{i \in \Irr} n_i P_i$ and denote by 
$\iota^\mu_k : P_k \rightarrow P$, $\mu = 1,\dots,n_k$, 
the embedding of $P_k$ into the $\mu$'th copy of $P_k$ in $P$.
Similarly, $r^\mu_k : P \rightarrow P_k$ is the projection onto
the $\mu$'th copy of $P_k$. Then
\be
  r^\mu_k \cir \iota^\nu_l = \delta_{k,l} \, \delta_{\mu,\nu} \,
	   \id_{P_k}
  \qquad \text{and} \qquad
  \sum_{k \in \Irr} \sum_{\mu=1}^{n_k} \iota^\mu_k \cir r^\mu_k 
	= \id_{P}  \ .
\labl{eq:P-into-Pk-prop}
Let $\beta$ be the conformal three-point block 
$P \times Q \times \Vc \rightarrow \Cb$ such that $d = b_\beta$.
According to \erf{eq:ev-gives-all} we can write
$\beta(p,q,v) = \ev_{\!P}\big(p,\beta^\sharp(q),v\big)$
with $\beta^\sharp$ an intertwiner from $Q$ to $P^*$. 
This in turn implies that 
${\underline d} = \bb_{\ev_{\!P}} \cir 
(\id_P \oti \bar \beta^\sharp)$, where (as usual) the bar over 
$\beta^\sharp$ indicates that it now acts on the right-movers. 
Thus we can define an intertwiner 
$\hat\varphi : P \oti \bar Q \rightarrow \hat\Hc$ (with $\hat\Hc$ as
given in (\ref{hpp})) by
\be
  \hat\varphi = \sum_{k \in \Irr} \sum_{\mu=1}^{n_k} 
  (r^\mu_k \oti (\bar \iota^\mu_k)^*) \cir
  (\id_P \oti \bar \beta^\sharp) \ . 
\ee
This intertwiner obeys
\begin{eqnarray}
\bb_{\ev} \cir \hat\varphi
& = & \displaystyle
\sum_{k,\mu} \bb_{\ev_{\!P_k}} \cir 
  (r_k^\mu \oti \bar \iota^{\mu_k\, *})
  \cir (\id_P \oti \bar \beta^\sharp) = 
\sum_{k,\mu} \bb_{\ev_{\!P}} \cir 
  (\id_P \oti (\bar r_k^{\mu_k\, *} \circ \bar \iota^{\mu_k\, *}))
  \cir (\id_P \oti \bar \beta^\sharp)
\nonumber \\
& = & \displaystyle
\sum_{k,\mu} \bb_{\ev_{\!P}} \cir 
(\id_P \oti (\bar \iota^\mu_k \cir \bar r^\mu_k)^*)
\cir (\id_P \oti \bar \beta^\sharp)
= \bb_{\ev_{\!P}} \cir (\id_P \oti \bar \beta^\sharp) 
= {\underline d} \ ,
\end{eqnarray}
where we used \erf{eq:move-g-in-ev} and \erf{eq:P-into-Pk-prop}.
Let $\varphi$ be the map from $P \oti \bar Q$
to the quotient $\Hc_\text{bulk} = \hat\Hc / \Nc_\ev$ induced
by $\hat\varphi$, {\it i.e.}\ $\varphi = \pi \cir \hat\varphi$, where 
$\pi : \hat\Hc \rightarrow \Hc_\text{bulk}$ is the projection to the
quotient. It then follows that for $x \in P \oti \bar Q$,
\be
  {\underline d}(x) = 0
  \quad\Rightarrow\quad
  \bb_\ev \cir \hat \varphi(x) = 0
  \quad\Rightarrow\quad
  \hat\varphi(x) \in \ker(\bb_\ev)
  \quad\Rightarrow\quad
  \varphi(x) = 0 ~.
\ee
Thus $\varphi$ vanishes on $K$, and since
$\Hc_1 \cong (P \oti \bar Q)/K$, $\varphi$ can be lifted to a
map starting at $\Hc_1$,
$\varphi' : \Hc_1 \rightarrow \Hc_\text{bulk}$.
For $x \in P \oti \bar Q$ and $[x]$ the corresponding class in
$\Hc_1$ we can write
\be
  \bb_\text{disc} \cir \varphi'([x])
  = \bb_\text{disc} \cir \varphi(x)
  = \bb_\text{disc} \cir \pi \cir \hat\varphi(x)
  = \bb_\ev \cir \hat\varphi(x)
  = {\underline d}(x)
  = \bb_1([x]) ~.
\ee
This shows that $\varphi'$ is an intertwiner such that
\erf{eq:maximal-universal} commutes. As mentioned below
\erf{eq:maximal-universal} this already implies that
$\varphi'$ is unique and injective.

\noindent Altogether we have therefore shown that
$(\Hc_\text{bulk},b_\text{disc})$ is indeed 
maximal.

\sect{The bulk space of the $c_{1,p}$ triplet models}\label{sec:Wp-minmod}

In this section we want to apply the abstract construction of the
previous section to the case of the $c_{1,p}$ triplet models. We begin by
collecting some basic properties of the representation theory of the 
$c_{1,p}$ triplet algebra $\Wc_p$.

\subsection{Representations of the $\Wc_p$-algebra}\label{sec:rep-Wp} 

The representation theory of the $\Wc_p$-algebra \cite{Kausch91}
has been analysed in 
\cite{Eholzer:1993ak,Flohr95,Kau95,Flohr:1996vc,GKau96b,%
Fuchs:2003yu,Carqueville:2005nu,Feigin:2005xs,Fuchs:2006nx}. Let us briefly
summarise the aspects we will need in the following.    

For a given $p \in \Zb_{\ge 2}$ the central charge of the
$\Wc_p$-algebra is $c=13-6p-6/p$. The $\Wc_p$-algebra has $2p$
irreducible representations that we shall label as
\be
U_s^\epsilon \ , \qquad s=1,\ldots, p \, \quad \epsilon = \pm \ .
\ee
Here $U_1^+$ is the vacuum
representation and $U_1^-$ describes the simple current.
For $p=2$, $U^-_1$ is
the representation $\Vc_1$ of \cite{GKau96b} and the 
irreducible representations $U_2^\pm$ are
$\Vc_{-1/8}$ and $\Vc_{3/8}$.

The projective cover of $U_s^\epsilon$ is denoted by  
$P_s^\epsilon$; for $s=p$, $U_p^\pm = P_p^\pm$, while for 
$s=1,\ldots p{-}1$ we have the maximal inclusions
\be
  U^\eps_s \subset M^\eps_{+,s} \subset N^\eps_s \subset P^\eps_s
  \qquad \text{and} \qquad
  U^\eps_s \subset M^\eps_{-,s} \subset N^\eps_s \subset P^\eps_s \ ,
\labl{eq:rep-include}
where furthermore $U^\eps_s = M^\eps_{+,s} \cap M^\eps_{-,s}$. Here 
$M^\eps_{\pm,s}$, $N^\eps_s$ and $P^\eps_s$ are indecomposable 
representations\footnote{
We follow the conventions used in \cite[sect.\,6]{Fuchs:2006nx}.
The relation to \cite[sect.\,2]{Fuchs:2003yu} is as follows:
$\Lambda(s) = U^+_s$,
$\Pi(s) = U^-_s$,
$\mathcal{R}_0(s) = P^+_s$,
$\mathcal{R}_1(s) = P^-_{p-s}$, as well as
$\mathcal{N}_0(s) =N^+_s$, 
$\mathcal{N}_1(s) = N^-_{p-s}$,
$\mathcal{N}_0^\pm(s) = M^+_{\pm,s}$ and
$\mathcal{N}_1^\pm(s) = M^-_{\pm,p-s}$.}.
The $\Wc_p$-algebra has an 
infinite number of distinct indecomposable
representations \cite{Feigin:2005xs}, but only the ones mentioned above
will be needed in our analysis.

For the case of the $c=-2$ triplet model at $p=2$, the two
indecomposable representations ${\cal R}_0$ and ${\cal R}_1$
of \cite{GKau96b} correspond to $P_1^\pm$. Furthermore, in the
symplectic fermion language (we are using the conventions of
\cite{Gaberdiel:2006pp}), the intermediate representations
$M^\pm_{\pm,1}$ and $N_1^\eps$ can be easily described: if we denote 
the subspace that is generated from $\chi^\pm_0\omega$ by  
the action of the fermionic modes by $M_{\pm,1}$ then
$M^+_{\pm,1}$ denotes the bosonic states in $M_{\pm,1}$, while
$M^-_{\pm,1}$ are the fermionic states. Both subspaces form then
representations of the (bosonic) triplet algebra. On the other hand, 
$N_1^\eps$ is the bosonic ($\eps = +$) or fermionic ($\eps = -$)
subspace generated by $\chi^+_0 \omega$ and $\chi^-_0 \omega$
together.

\subsection{Intertwiners}\label{sec:intertwiners}

For the following it is important to understand the space of
intertwiners between two (indecomposable) projective representations 
$P^\eps_s$ and $P^\nu_t$. To this end we consider the exact sequences 
(see appendix \ref{app:fusion})
\be
  0 \rightarrow N^\eps_s \rightarrow P^\eps_{s} \rightarrow U^{\eps}_{s}
  \rightarrow 0 \ ,\qquad
  0 \rightarrow M^\eps_{\nu,s} \rightarrow P^\eps_{s} \rightarrow 
  M^{-\eps}_{\nu,p-s} \rightarrow 0 \ ,
\labl{eq:exact-P-middle}
where $s=1,\dots,p{-}1$ and $\eps,\nu=\pm$. 
Together with \erf{eq:rep-include} it follows that
$\Hom_\Vc(P^\eps_s,P^\eps_s)$ contains at least two linearly
independent maps, namely the identity $\id$, and 
\be
n : P^\eps_{s} \rightarrow U^{\eps}_{s}\rightarrow P^\eps_{s} \ , 
\labl{eq:n-def}
where the intermediate maps are the surjection of the projective cover
and the embedding $U^\eps_s \subset P^\eps_s$.
In the symplectic fermion language for $p\,{=}\,2$, 
the intertwiner $n$ is simply $n=\chi^-_0 \, \chi^+_0$.

Similarly, the intertwiners $\Hom_\Vc(P^\eps_s,P^{-\eps}_{p-s})$
contain (we are suppressing the dependence on $s$ and $\eps$ in the
definition of $e_\nu$)
\be
e_\nu: P^\eps_{s} \rightarrow  
  M^{-\eps}_{\nu,p-s} \rightarrow P^{-\eps}_{p-s} \qquad
\nu = \pm\ ,
\labl{eq:e_mu-def}
where the intermediate maps are those appearing in \erf{eq:exact-P-middle}.
Again for $p=2$, we simply have $e_\pm = \chi^\pm_0$. It is argued in
appendix \ref{app:fusion} that
the identity map id, \erf{eq:n-def} and \erf{eq:e_mu-def}
already give all intertwiners,  
\be\label{interitw}
  \Hom_\Vc(P^\eps_s,P^\nu_t) = \begin{cases} 
  \Cb\, \id \oplus \Cb\, n &;~ t\,{=}\,s~,~\nu\,{=}\,\eps \\
  \Cb \, e_+ \oplus \Cb\, e_- &;~ t\,{=}\,p{-}s~,~\nu\,{=}\,{-}\eps \\
  \{0 \} &;~ \text{otherwise.}
  \end{cases}
\ee

The dimension of the Hom-spaces can also be understood from the following
diagrams which describe the sub-quotients of $P^+_s$ and $P^-_{p-s}$,
\be
P^+_s \,:\,
\raisebox{-2.2em}{
\begin{xy}
(10,20)*+{U^+_s}="a"; %
(0,10)*+{U^-_{p-s}}="b"; %
(20,10)*+{U^-_{p-s}}="c"; %
(10,0)*+{U^+_s}="d"; %
{\ar@{->} "a";"b"}; %
{\ar@{->} "a";"c"}; %
{\ar@{->} "b";"d"}; %
{\ar@{->} "c";"d"}; %
\end{xy}}
\ , \qquad
P^-_{p-s} \,:\,
\raisebox{-2.2em}{
\begin{xy}
(10,20)*+{U^-_{p-s}}="a"; %
(0,10)*+{U^+_s}="b"; %
(20,10)*+{U^+_s}="c"; %
(10,0)*+{U^-_{p-s}}="d"; %
{\ar@{->} "a";"b"}; %
{\ar@{->} "a";"c"}; %
{\ar@{->} "b";"d"}; %
{\ar@{->} "c";"d"}; %
\end{xy}}
~~.
\labl{eq:P+P-pic}
The entries in the first diagram correspond to the inclusions
\erf{eq:rep-include} in the sense that we have
the equivalences $P^+_s / N^+_s \cong U^+_s$ and
$N^+_s / U^+_s \cong U^-_{p-s} \oplus U^-_{p-s}$. The arrows
indicate that {\it e.g.}\ a representative $v$ of a non-zero
$[v] \in P^+_s / N^+_s$ (the copy of $U^+_s$ at the top
of the diagram) can get mapped to $N^+_s$, but not
vice versa.

The exact sequences \erf{eq:exact-P-middle} together with the
inclusions \erf{eq:rep-include} also imply that, for $\nu = \pm$, 
\be\label{nn}
  n \cir n = 0 \ ,\qquad
  e_\nu \cir n = 0 \ , \qquad
  n \cir e_\nu = 0  \ , \qquad
  e_\nu \cir e_\nu = 0 \ .
\ee
The combination $e_- \cir e_+$ acting on, say, $P^+_s$ amounts to the
composition 
\be
P^+_s \twoheadrightarrow M^-_{+,p-s} \hookrightarrow P^-_{p-s}
\twoheadrightarrow M^+_{-,s} \hookrightarrow P^+_{s}\ . 
\ee
The kernel of the second surjection is $M^-_{-,p-s}$, so that elements
of $P^+_s$ which get mapped to elements of $M^-_{+,p-s}$ that do not
also lie in $M^-_{-,p-s}$, have a non-zero image in $P^+_s$. In
particular, $e_- \cir e_+ \neq 0$. Since $e_- \cir (e_- \cir e_+) = 0$
we see that $e_- \cir e_+$ is proportional to $n$. Similarly one can
check that $e_+ \cir e_-$ is nonzero and proportional to $n$. We
choose to normalise $n$ such that 
\be
  e_- \cir e_+ = n
  \ , \qquad 
  e_+ \cir e_- = \lambda n
  \qquad \text{for some}~~\lambda \in \Cb^\times \ .
\ee
For the case of $p=2$, it follows from the above identifications that
$\lambda=-1$. In general $\lambda$ may also depend on $s$ and $\eps$.

\subsection{The construction of the bulk space}\label{sec:bulk-construct}

We now want to apply the general construction of 
section~\ref{sec:identity-brane} to the case
of the $c_{1,p}$ triplet models. We shall assume that the theory has a
boundary condition for which $\Hc_{\text{bnd}}=\Vc$. This is motivated
by the analysis of \cite{Gaberdiel:2006pp} where for the $p=2$ example
the boundary conditions were analysed in detail (using the symplectic 
fermion description). For $p=2$ we found that the boundary conditions 
are labelled by the irreducible representations and that the open
string spectrum is simply determined by the corresponding fusion
rules. This led to the conjecture that the same structure is also
present for the other $c_{1,p}$ triplet models. If this is the case, then
the brane associated to the irreducible vacuum representation has
indeed  $\Hc_{\text{bnd}}=\Vc$. 

We shall thus assume that such a brane exists and determine the
corresponding bulk spectrum following the strategy of 
section~\ref{sec:identity-brane}. As a
consistency check we shall later study the boundary states of the
resulting bulk theory (see section~\ref{sec:boundary-states}). 
We shall find that our bulk 
theory has indeed boundary states associated to the irreducible
representations of the triplet algebra, and that their open string
spectra are determined by the fusion rules. This therefore forms a
stringent consistency check on our procedure.
\medskip

With this in mind, all we have to do is to calculate the kernel  
$\Nc_\ev$ of the map $\bb_{\ev}$. Recall from section \ref{sec:kernel-bev}
the definition of the spaces $H_{kl}$, as
well as the maps $c_{kl}$ and $d_{kl}$.
It follows from
(\ref{interitw}) that the spaces $H_{kl}$ are  only
nonzero for  $(k,l) = \big((s,\eps),(s,\eps)\big)$ with $s=1,\ldots,p$
and $(k,l) = \big((s,\eps),(p{-}s,-\eps)\big)$ for $s=1,\ldots,p{-}1$,
where in both cases $\eps=\pm$. 
It is therefore enough to consider the
kernel of $\bb_\ev$ separately for the spaces 
\be
\hat\Hc_s = (P^+_s \oti \bar P^{+\,*}_s )\oplus 
            (P^-_{p-s} \oti \bar P^{-\,*}_{p-s})\ ,
\ee
where $s=1,\ldots,p{-}1$. (Since $P_p^\pm$ is already 
irreducible, the kernel in the summands with $s=p$ is trivial.)
The spaces $H_{kl}$ are of the form $H^+ := H_{(s,\eps)(s,\eps)}$ and 
$H^- := H_{(s,\eps)(p-s,-\eps)}$, where
\be\bearll
  H^+ \etb\!\!= 
  \Hom_\Vc(P^\eps_s,P^\eps_s) \oti \Hom_\Vc(P^\eps_s,P^\eps_s) ~\oplus~ 
  \Hom_\Vc(P^\eps_s,P^{-\eps}_{p-s}) \oti 
\Hom_\Vc(P^{-\eps}_{p-s},P^\eps_s) \ ,
  \enl
  H^- \etb\!\!= 
  \Hom_\Vc(P^\eps_s,P^\eps_s) \oti \Hom_\Vc(P^\eps_s,P^{-\eps}_{p-s}) 
  ~\oplus~ 
\Hom_\Vc(P^\eps_s,P^{-\eps}_{p-s}) \oti 
\Hom_\Vc(P^{-\eps}_{p-s},P^{-\eps}_{p-s}) \ .
\eear\ee
An element $u$ of $H^+$ is a linear combination of the form
\be
  u = a \,(\id \oti \id) + b \, (n \oti \id) + c \, (\id \oti n) 
     + d \, (n \oti n) +
  \sum_{\mu,\nu = \pm} f^{\mu\nu} \, (e_\mu \oti e_\nu) \ .
\ee
Applying $c^+ := c_{(s,\eps)(s,\eps)}$ to $u$ yields
$c^+(u) = a \, \id + (b{+}c{+}f^{+-} + \lambda f^{-+}) \, n$. The
kernel of $c^+$ is thus given by 
\bea
  \ker(c^+) = \text{span}_\Cb\big\{~
  (n \oti n) \,,~ (e_+ \oti e_+) \,,~  (e_- \oti e_- )\,,~ 
  (n \oti \id - \id \oti n) \,,~
  \\
  \hspace*{10em}
  (\lambda \,e_+ \oti e_- - e_- \oti e_+) \,,~
  (n \oti \id - e_+ \oti e_-) ~\big\} \ .
\eear\labl{eq:Wp-ker-plus}
Similarly, an element $v$ of $H^-$ is a linear combination of the form
\be
  v = \sum_{\nu = \pm}\Big( 
  a^\nu \,(\id \oti e_\nu) + b^\nu \, (e_\nu \oti \id) + 
  c^\nu \,(n \oti e_\nu) + d^\nu \, (e_\nu \oti n) \Big) \ .
\ee
Applying $c^- := c_{(s,\eps)(p{-}s,-\eps)}$ to $v$ gives
$c^-(v) = \sum_\nu (a^\nu + b^\nu) e_\nu$, so that
\be
 \ker(c^-) = \text{span}_\Cb\big\{~
  (n \oti e_\nu)\,,~(e_\nu \oti n)\,,~
  (\id \oti e_\nu - e_\nu \oti \id) \,\big|\,\nu = \pm ~\big\} \ .
\labl{eq:Wp-ker-minus}
Using (\ref{eq:b1-kernel-general}) the kernel $\Nc_\ev$ of $\bb_{\ev}$ is
now simply $d^\pm(f)$ for the various generators $f$ of $\ker(c^\pm)$.

\medskip

This specifies the kernel $\Nc_\ev$ completely. However,
we do not need to consider the image of all of these maps
separately. In fact, the kernel is already generated by the images of
the last element of $\ker(c^-)$ with $\nu=\pm$. More precisely, we
define the space
\be
{\cal D}_s = 
(P^+_s \oti \bar P^{-\, *}_{p-s}) 
~\oplus~ (P^-_{p-s} \oti \bar P^{+\, *}_{s}) 
 \ . 
\ee
Then 
\be
{\cal K }^\nu_s =  
(\id \oti \bar e_\nu^* - e_\nu \oti \bar \id) \, {\cal D}_s \ , \qquad
\nu = \pm 
\ee
are subspaces of $\hat{\Hc}_s$. (Recall that $e_\nu$ maps $P_s^+$ to
$P_{p-s}^-$ and vice versa, and it is understood that 
$(\id \oti \bar e_\nu^* - e_\nu \oti \bar \id)$ 
acts on both summands of ${\cal D}_s$.) We now claim that the 
kernel of $\bb_{\ev}$ restricted to $\hat\Hc_s$
is simply the span of these two spaces,  
\be\label{Ns} 
\Nc_s = \ker\big( \bb_{\ev}\big|_{\hat\Hc_s} \big)
= {\rm span}_{\Cb} \Bigl({\cal K}^{+}_s ,~ {\cal K}^{-}_s 
\Bigr) \ . 
\ee
By construction, it is clear that ${\cal K}_s^{\pm}\subset \Nc_s$;
it only remains to prove that they generate already all of $\Nc_s$,
{\it i.e.}\ that the images of $d^\pm(f)$ for the
various generators $f$ of $\ker(c^\pm)$ lie in a linear combination of
states from ${\cal K}_s^{\pm}$. Let us check this explicitly in two 
examples; the rest can be seen similarly.     
To obtain the first generator of $\ker(c^+)$ in
\erf{eq:Wp-ker-plus} we consider the composition
\be
  (\id \oti \bar e_+^* - e_+ \oti \bar \id) \cir (n \oti \bar e_-^*)
  = n \oti (\bar e_+^* \cir \bar e_-^*) 
           - (e_+ \cir n) \oti \bar e_-^*
  = n \oti (\bar e_- \cir \bar e_+)^* = n \oti \bar n^* \ .
\ee
Similarly, the forth generator in \erf{eq:Wp-ker-plus} is obtained by
taking  
\be\begin{array}{rl}
& - (\id \oti \bar e_+^* - e_+ \oti \bar\id) \cir (\id \oti \bar e_-^*)
  - (\id \oti \bar e_-^* - e_- \oti \bar\id) \cir (e_+ \oti \bar \id)
  \\[.2em]
  &= -\id \oti (\bar e_+^* \cir \bar e_-^*) + e_+ \oti \bar e_-^*
  - e_+ \oti \bar e_-^* + (e_- \cir e_+) \oti \bar \id 
  = - \id \oti \bar n^*  + n \oti \bar \id \ .
\eear\ee

\subsection{The bulk space and comparison to $p=2$}
\label{sec:W2-bulk-space}

Summarising the above discussion we therefore find that the 
actual bulk space of states $\Hc_{\text{bulk}}$ is  of the form 
\be
\Hc_{\text{bulk}} = \bigoplus_{s=1}^{p-1} 
\hat{\Hc}_s / \Nc_s   ~ \oplus ~
  (U^+_p \oti \bar U^{+\,*}_p) ~ \oplus ~
  (U^-_p \oti \bar U^{-\,*}_p)  \ , 
\labl{eq:Wp-bulk-space}
where $\Nc_s$ is defined in (\ref{Ns}). 
For $p=2$ eqn.~\erf{eq:Wp-bulk-space} becomes in
the notation of \cite{Gaberdiel:2006pp}
\be
\Hc_{\text{bulk}} = {\cal H}^{bos}_{\omega} / \Nc \oplus
\Bigl( \Vc_{-1/8}\oti \bar\Vc_{-1/8} \Bigr) ~\oplus~
\Bigl( \Vc_{3/8}\oti \bar\Vc_{3/8} \Bigr) \ ,
\ee
where ${\cal H}^{bos}_{\omega} = ({\cal R}_0 \oti \bar{\cal R}_0) 
\oplus ({\cal R}_1 \oti \bar{\cal R}_1)$. Furthermore, it follows from
(\ref{Ns}) together with the identification of $e_\pm=\chi^\pm_0$ 
that $\Nc$ consists of the states
\be
\Nc = {\rm span}_{\Cb} \big\{
(\chi^\nu_0 - \bar\chi^\nu_0 )\, \psi ~\big|~ \nu=\pm \ , ~ 
\psi \in {\cal H}^{fer}_{\omega} \big\} \ ,
\ee
where ${\cal H}^{fer}_{\omega} = ({\cal R}_0 \oti \bar{\cal R}_1) 
\oplus ({\cal R}_1 \oti \bar{\cal R}_0)$. This then reproduces
precisely the description of the bulk theory used in 
\cite{Gaberdiel:2006pp}.
\smallskip

In \cite{Quella:2007hr} harmonic analysis on supergroups 
was used to obtain the space of bulk states for WZW models with
supergroup targets. The similarities in the representation theory
of super Lie algebras and the $\Wc_p$-algebra were then exploited to 
propose a description of the bulk space as 
\be\label{QSbulk}
\Hc_{\text{bulk}} = \bigoplus_{s=1}^{p-1} 
{\cal I}_s  ~ \oplus ~
  (U^+_p \oti \bar U^{+\,*}_p) ~ \oplus ~
  (U^-_p \oti \bar U^{-\,*}_p)  \ ,
\ee
and a composition series for ${\cal I}_s$ was given. (However, 
unlike the expression \erf{eq:Wp-bulk-space} in terms of quotients, a
composition series does in general not fix a representation up
to isomorphism.) In \cite{Quella:2007hr} it is also conjectured that
as a $\bar\Wc_p$-representation (but not as a 
$\Wc_p{\times}\bar\Wc_p$-representation)
$\mathcal{I}_s$ is of the form
\be
{\cal I}_s = (U_s^+ \oti \bar P^{+\,*}_s)
\oplus (U^-_{p-s} \oti \bar P^{-\,*}_{p-s}) \ .
\labl{decomp1}
In order to compare this prediction with our result we now have to
decompose our quotient space $\Hc_s \equiv \hat\Hc_s / \Nc_s$ with
respect to the $\bar\Wc_p$ action. This is done in
appendix~\ref{app:proj-inj-ker} and we find agreement with 
\erf{decomp1}. It is encouraging that the two
proposals fit together.

\subsection{Modular invariance for the $c_{1,p}$ triplet models}
\label{sec:Wp-mod-inv}

Finally, we want to show that the partition function of the bulk
space $\Hc_{\text{bulk}}$ is modular invariant. 
We have already proven in section \ref{sec:bulk-proj} that
the partition function is invariant under $\tau\mapsto \tau +1$. This
followed from the fact that 
$\exp( 2 \pi i(L_0{-}\bar L_0))$ acts as the identity on 
$\Hc_\text{bulk}$. In the present context this can be seen more
concretely because the element $n \oti \id - \id \oti n$ is in 
$\ker(c^+)$, see  eqn.\,\erf{eq:Wp-ker-plus}, and since 
$e^{2 \pi i L_0}\in \Hom_\Vc(P^\eps_s,P^\eps_s)$ 
can be written as a linear combination of $\id$ and $n$. 

With the help of equation \erf{decomp1} it is now straightforward to
compute the partition function of 
$\Hc_\text{bulk}$ in \erf{eq:Wp-bulk-space}. We will start by
recalling the expressions for the characters of the
$\Wc_p$-representations and their modular properties. These were
first described in \cite{Flohr95}; here we will follow the   
presentation in \cite[sect.\,3]{Fuchs:2003yu}. The characters of the  
irreducible representations are, for $s=1,\dots,p$ 
and $\nu\,{=}\,\pm$,
\be
   \chi_{U^+_s}(\tau) = \frac{1}{\eta(q)} \left(
   \frac{s}{p}\theta_{p-s,p}(q) +  2 \theta'_{p-s,p}(q) \right) \ ,
~~
\chi_{U^-_s}(\tau) = \frac{1}{\eta(q)} \left(
   \frac{s}{p}\theta_{s,p}(q) -  2 \theta'_{s,p}(q) \right) \ .
\ee
Here $\eta(q)$ is the Dedekind eta function and
$\theta_{s,p}(q)=\theta_{s,p}(1,q)$ with
\be
\eta(q) = q^{\frac{1}{24}} \prod_{n=1}^\infty (1-q^n)\ , \qquad 
\theta_{s,p}(z,q) = \sum_{m\in \Zb+\tfrac{s}{2p}} z^m q^{pm^2}\ .
\ee
We also define
\be
\theta_{s,p}'(q)= z \left. \frac{\partial}{\partial z} 
\theta_{s,p}(z,q) \right|_{z=1}\ . 
\ee
Both $\theta_{s,p}$ and $\theta'_{s,p}$ are periodic in $s$ with
period $2p$. In addition we have 
$\theta_{s,p}(q) = \theta_{2p-s,p}(q)$ and 
$\theta'_{s,p}(q) = - \theta'_{2p-s,p}(q)$. Thus we can restrict $s$
to take the values $s=0,\ldots, p$. Furthermore it follows that 
$\theta'_{0,p}(q) =  0 = \theta'_{p,p}(q)$. The modular
transformation properties under $\tau\mapsto -1/\tau$ are 
\begin{eqnarray}
\frac{\theta_{s,p}}{\eta} \left(-\frac{1}{\tau}\right) & = &  
\frac{1}{\sqrt{2p}} \, 
\sum_{s'=0}^{2p-1} e^{i\pi s s' / p} \, \frac{\theta_{s',p}}{\eta}(\tau)
\nonumber \\
& = & \frac{1}{\sqrt{2p}} \, \left[
\frac{\theta_{0,p}}{\eta}(\tau) + 
(-1)^s \, \frac{\theta_{p,p}}{\eta}(\tau)
+ 2 \sum_{s'=1}^{p-1}  \cos\!\left(\frac{\pi s s'}{p}\right) \, 
\frac{\theta_{s',p}}{\eta}(\tau) \right] \ , \label{thet}
\end{eqnarray}
and similarly
\be
\frac{\theta'_{s,p}}{\eta} \left(-\frac{1}{\tau}\right)  =   
-\frac{\tau }{\sqrt{2p}} \, 
\sum_{s'=0}^{2p-1} e^{i\pi s s' / p} \, 
\frac{\theta'_{s',p}}{\eta}(\tau)
 =  - \frac{2i\tau}{\sqrt{2p}} \,
\sum_{s'=1}^{p-1}  \sin\!\left(\frac{\pi s s'}{p}\right) \, 
\frac{\theta'_{s',p}}{\eta}(\tau)  \ . 
\labl{thetp}

\noindent For the following it is also useful to 
abbreviate, for $s=1,\dots,p{-} 1$, 
\be
\psi^+_s(\tau) = \chi_{U^+_s}(\tau) + \chi_{U^-_{p-s}}(\tau)
   = \frac{\theta_{p-s,p}(q)}{\eta(q)}
   \ , \quad
\psi^-_s(\tau) = \chi_{U^-_s}(\tau) + \chi_{U^+_{p-s}}(\tau)
   = \frac{\theta_{s,p}(q)}{\eta(q)} \ .
\labl{eq:psi+-def}
It then follows from the exact sequences in 
section~\ref{sec:intertwiners} that
$\text{tr}_{P^\pm_s} \big( q^{L_0-c/24} \big) = 2 \psi^\pm_s(\tau)$
for $s=1,\ldots,p{-}1$.
Equation \erf{decomp1} now implies that \
\be
   \text{tr}_{\Hc_s}
   \big( q^{L_0-c/24} (q^*)^{\bar L_0 - c/24}  \big) 
   = 2 \chi_{U^+_s}(\tau) \overline{\psi_s^+(\tau)} 
     + 2 \chi_{U^-_{p-s}}(\tau) \overline{\psi_{p-s}^-(\tau)} 
   = 2 |\psi_s^+(\tau)|^2  \ ,
\ee
where we have used that $\psi_{p-s}^-=\psi_s^+$.
Taking the trace of $\Hc_\text{bulk}$ and using the previously
mentioned relations between the different $\theta_{s,p}$ functions
then gives
\be\bearll
   Z(\tau) \etb\!\!=~
\text{tr}_{\Hc_\text{bulk}}
\big( q^{L_0-c/24} (q^*)^{\bar L_0 - c/24} \big)
\\ \displaystyle
\etb\!\!=~ |\chi_{U^+_p}(\tau)|^2 + |\chi_{U^-_p}(\tau)|^2 +
    2 \sum_{s=1}^{p-1} |\psi_s^+(\tau)|^2
~=~ \frac{1}{|\eta(\tau)|^2}
\sum_{s=0}^{2p-1} | \theta_{s,p}(q) |^2 ~.
\eear\ee
As already noted in \cite{Flohr95} (see also \cite{Quella:2007hr}), 
the last expression for $Z(\tau)$ is easily checked to obey
$Z(-1/\tau) = Z(\tau)$ using the modular properties   
of the theta functions.

\sect{Boundary states}\label{sec:boundary-states}

Now that we have the bulk spectrum under control we can analyse the
possible Ishibashi states and construct the boundary
states. This will be a consistency check of our approach since we
started out by assuming that the theory possesses an `identity brane'
whose open string spectrum only consists of the chiral algebra itself.

\subsection{The space of Ishibashi states}\label{sec:Ishi-space}

Boundary conditions are usually described in terms of boundary states
that live in a suitable completion of $\Hc_\text{bulk}$. Instead of
giving the boundary state explicitly, we may also specify it by giving
all bulk one-point functions on the disc (or on the upper half
plane). Thus we may think of a boundary condition as being described
by a linear form $\Hc_\text{bulk} \rightarrow \Cb$. For concreteness,
we take this linear form to be defined by inserting the bulk field on
the upper half plane at the point $z=i$.

Every boundary condition that preserves the symmetry described by
$\Vc$ necessarily contains $\Vc$ as a subspace of $\Hc_{\text{bnd}}$, 
and we may as well consider the correlator of a bulk field inserted
at $z=i$, together with a boundary field in the subspace $\Vc$ inserted
at $0$. Thus we are led to consider a bilinear map
$\Hc_\text{bulk} \ti \Vc \rightarrow \Cb$. The compatibility
conditions with the action of $\Vc$ for such a bilinear map are just
the same as for $b_\text{disc}$. After all, $b_\text{disc}$ is
precisely such a correlator for the identity brane.
The space of bilinear maps $\Hc_\text{bulk} \ti \Vc \rightarrow \Cb$
compatible with the action of $\Vc$ is by definition the space of
Ishibashi states, and we therefore obtain\footnote{
The term Ishibashi state is more commonly
used for a state in $\Hc_\text{bulk}$ and appears in the
description of boundaries without insertions of $\Vc$.
The two descriptions are equivalent. For example, given a
$b \in B(\Hc_\text{bulk})$,
the corresponding element $\phi \In \Hc_\text{bulk}$
is determined by
$b(\phi',\Omega) = \big\langle \phi(-i) \phi'(i)\big\rangle$
for all $\phi' \In \Hc_\text{bulk}$. Here
the right hand side is the bulk two point function on the
complex plane and
$\Omega \in \Vc$ is the vacuum vector of the vertex algebra.
(To obtain boundary states for the disc, rather than the
upper half plane, one should employ an appropriate conformal
transformation.)}
\be
  B(\Hc_\text{bulk}) ~~
  \text{is the space of Ishibashi states for} ~~
  \Hc_\text{bulk} ~.
\labl{eq:Ishibashi-form1}
Here $\Hc_\text{bulk}$ is the quotient \erf{eq:bulk-general} and
$B(\Hc_\text{bulk})$ is defined as in
section \ref{sec:universal} and appendix \ref{app:b-mode-com}.

Since $\Hc_\text{bulk}$ is defined as the quotient
$\hat\Hc/\Nc_\ev$, $B(\Hc_\text{bulk})$
is isomorphic to the space of linear forms on $\hat\Hc$ that lie in
$B(\hat\Hc)$ and that vanish on $\Nc_\ev$, 
\be
B(\Hc_{\text{bulk}}) \cong 
\big\{\, b \in B(\hat\Hc) \, \big| \, b(v)\,{=}\,0~
\text{for all}~ v \in \Nc_\ev \,\big\} \ .
\ee
Because of \erf{eq:ev-gives-all} and \erf{eq:move-g-in-ev} every
element $b\in B(\hat\Hc)$ can be written as 
\be
b(p \oti \bar q,v)
= \sum_{k \in \Irr} \ev_{P_k}(\rho_k(p),q,v) 
\ee
for an appropriate $\rho_k \in \Hom_\Vc(P_k,P_k)$. In fact, this
defines an isomorphism
\be
\bigoplus_{k \in \Irr} \Hom_\Vc(P_k,P_k)
  \overset{\cong}{\longrightarrow}
  B(\hat\Hc) \ .
\ee
By the same arguments used to obtain the description of
$\Nc_\ev$ in \erf{eq:b1-kernel-general},
it is not hard to see that $\bb$
vanishes on $\Nc_\ev$
if and only if $\sum_i g_i \cir \rho_i \cir f_i = 0$ whenever
$\sum_i g_i \cir f_i = 0$, {\it i.e.}\ whenever 
$\sum_i g_i \oti f_i$ lies in the kernel of $c_{kl}$ as defined
in \erf{eq:cdef}. But let us see more concretely what the result is 
for the $c_{1,p}$ triplet models.
\medskip

For the case of the $c_{1,p}$ triplet models, 
we can consider the various
summands of $\Hc_{\text{bulk}}$ in \erf{eq:Wp-bulk-space} 
separately. 
The two irreducible summands corresponding to $P_p^\pm$ give rise to 
two Ishibashi  
states corresponding to $\rho=\id_{P^+_p}$ and $\rho=\id_{P^-_p}$,
respectively. (These are the familiar Ishibashi states associated to
irreducible representations.) The situation is more interesting for
$\Hc_s=\hat\Hc_s/\Nc_s$ with $s=1,\ldots,p{-}1$. The space of
intertwiners  
\be
(\rho_+,\rho_-)\in 
\Hom_\Vc(P_s^+,P_s^+) \oplus \Hom_\Vc(P_{p-s}^-,P_{p-s}^-)
\ee
is $4$-dimensional, but we also have to impose the condition that 
$(\rho_+,\rho_-)$ vanishes on $\Nc_s$. Since $\Nc_s$ is generated by 
${\cal K}_s^\pm$, we thus need to analyse whether 
$(\rho_+,\rho_-)$ vanishes on ${\cal K}_s^\pm$. Using
(\ref{eq:move-g-in-ev}) this leads to the condition
\be\begin{array}{ll}\displaystyle
e_\nu \circ \rho_+ - \rho_- \circ e_\nu = 0 \etb \in \Hom_\Vc(P^+_s,P^-_{p-s})
\ , \quad
\hbox{and} \\[.2em] \displaystyle
e_\nu \circ \rho_- - \rho_+ \circ e_\nu = 0 \etb \in \Hom_\Vc(P^-_{p-s},P^+_s)
\ ,
\eear\ee
where $\nu=\pm$. A short calculation using \erf{nn} shows that the
space of solutions is three dimensional and is given by
\be
\rho_+ = \alpha \, \id + \beta \, n \ , \qquad
\rho_- = \alpha \, \id + \gamma \, n \ ,
\ee
where $\alpha,\beta,\gamma \in \Cb$.
(In the first equation $\id$ and $n$ act on $P^+_s$, while 
in the second they act on $P^-_{p-s}$.)
Altogether we therefore obtain $3(p{-}1)$
Ishibashi states from the indecomposable sectors $\Hc_s$, as well as
two Ishibashi states from the irreducible representations, giving in
total $3p-1$ Ishibashi states. Obviously this agrees with the explicit
analysis for $p=2$ in \cite{Gaberdiel:2006pp}. It also agrees with the
number of chiral torus amplitudes \cite{Flohr:2005cm}, as may have
been expected.

As in the case for $p=2$ we do not expect that all of these Ishibashi
states will contribute to the boundary states. Indeed, the space of
torus amplitudes contains only $2p$ functions that are power series in
$q$, while the remaining $p{-}1$ torus functions involve terms
proportional to $\tau$ \cite{Flohr:2005cm} (see also \cite{Flohr95}).
The latter cannot appear in a consistent open string expansion (since
the open string description involves a trace that can never lead to a
term proportional to $\tau$), and thus the space of open string
amplitudes is only $2p$ dimensional. But then it follows that only a 
$2p$ dimensional subspace of the Ishibashi states can contribute to
consistent boundary states. In fact, one would expect that for each 
$\Hc_s$ only two linear combinations of the three Ishibashi states can
contribute. This expectation is borne out by the detailed construction
to which we now turn.

\subsection{Constructing the boundary states}

The analysis of \cite{Gaberdiel:2006pp} suggests that the boundary states
of the `charge-conjugation' $c_{1,p}$ triplet models 
are labelled by the irreducible representations of the
$\Wc_p$-algebra. 
As we have explained before, the irreducible representations
are labelled by $(s,\eps)$, where $s=1,\ldots, p$ and $\eps=\pm$; we
shall denote the corresponding boundary states as
$|\!|(s,\eps)\rangle\!\rangle$. Given the results of  
\cite{Gaberdiel:2006pp} it is furthermore natural to expect
that their open string spectrum is described
by the fusion rules, {\it i.e.}\ that
\be
\langle\!\langle (s_1,\eps_1) |\!| 
q^{\frac{1}{2}(L_0+\bar{L}_0) -  \frac{c}{24}} 
|\!| (s_2,\eps_2) \rangle\!\rangle 
= \sum_{\cal R} \Nc^{\text{fus}}_{(s_1,\eps_1)\, 
(s_2,\eps_2)}{}^{{\cal R}} \,  
\text{tr}_{\cal R} (\tilde{q}^{L_0 - c/24}) \ , 
\labl{conj}
where as always in the following $q=e^{2\pi i \tau}$ and 
$\tilde{q}= e^{-2\pi i /\tau}$. Here 
$\Nc^{\text{fus}}_{(s_1,\eps_1)\, (s_2,\eps_2)}{}^{\cal R}$ gives the 
decomposition of the fusion product of two irreducible representations,
\be
  U^{\eps_1}_{s_1} \circledast U^{\eps_2}_{s_2} 
  = \bigoplus_{\mathcal{R}} 
  \Nc^{\text{fus}}_{(s_1,\eps_1)\, (s_2,\eps_2)}{}^{\cal R}
  \; \mathcal{R} ~,
\ee
for more details see appendix~\ref{app:fusion}. 
The ansatz (\ref{conj}) is the natural generalisation of the usual
Cardy situation \cite{Cardy:1989ir}
to rational logarithmic conformal field theories. 

On the level of characters one cannot tell the difference between
$P^\nu_s$ and $2U^\nu_s \oplus 2U^{-\nu}_{p-s}$, so that we may as
well write \erf{conj} directly in terms of the structure constants
$N$ of the Grothendieck ring, also given in appendix~\ref{app:fusion},
\be
\langle\!\langle (s_1,\eps_1) |\!| 
q^{\frac{1}{2}(L_0+\bar{L}_0) -  \frac{c}{24}} 
|\!| (s_2,\eps_2) \rangle\!\rangle 
= \sum_{r=1}^p \sum_{\mu = \pm} 
  N_{(s_1,\eps_1)(s_2,\eps_2)}^{\hspace{3em}(r,\mu)} 
  \,\chi_{U^\mu_r}(\tilde q) ~.
\labl{eq:conj-groth}

Starting from the ansatz \erf{eq:conj-groth}
we now want to construct the boundary
states explicitly. To this end, we will evaluate
\erf{eq:conj-groth} in the special case $(s_1,\eps_1)=(1,+)$
and $(s_2,\eps_2)$ arbitrary, as well as for
$(s_1,\eps_1)=(s_2,\eps_2)=(p,+)$. This will determine the
boundary states $|\!| (s,\eps) \rangle\!\rangle$ and
the overlap of the Ishibashi states. It is then a highly
non-trivial consistency check that all other overlaps
also agree with \erf{eq:conj-groth}.

\medskip

To carry out this calculation it is convenient to
work with the $S$-matrices of the $\Wc_p$-characters. These
can be found from the transformation properties of the
theta-functions in section \ref{sec:Wp-mod-inv} or by rearranging 
\cite[prop.\,3.4]{Fuchs:2003yu},
\be
  \chi_{U^\eps_s}(\tilde q)
  = \sum_{t=1}^{p} \sum_{\nu = \pm}
  \big( S_{(s,\eps),(t,\nu)} - i \tau S^\ell_{(s,\eps),(t,\nu)} \big) 
  \chi_{U^\nu_t}(q) \ .
\labl{eq:mod-xfer-S}
Here $\tilde{q}=e^{-2\pi i / \tau}$ is the open string loop parameter,
while $q=e^{2\pi i \tau}$ is the corresponding parameter in the closed
string. The matrices $S$ and $S^\ell$ are,
for $s,t = 1,\dots,p$ and $\eps,\nu = \pm$,
\bea
  S_{(s,\eps),(t,\nu)}
  = \frac{2 - \delta_{t,p}}{\sqrt{2p}}
  \, \frac{s}{p} \,
  \cos\!\Big( \pi \frac{s t}{p} \Big)
  (-\nu)^s
  (-\eps)^t
  (-1)^{p(\eps+1)(\nu+1)/4} \ ,
  \enl
  S^\ell_{(s,\eps),(t,\nu)}
  = \frac{2}{\sqrt{2p}}
  \, \frac{p-t}{p} \,
  \sin\!\Big( \pi \frac{s t}{p} \Big)
  (-\nu)^s
  (-\eps)^t
  (-1)^{p(\eps+1)(\nu+1)/4} \ .
\eear\labl{eq:mod-S-mat}
Note that $S^\ell_{(s,\eps),(p,\nu)} = 0$ and that  
for $t=1,\dots,p{-}1$ these $S$-matrices have the
symmetries
\be
S_{(s,\eps),(p-t,-\nu)}
= S_{(s,\eps),(t,\nu)} \ ,
\qquad
(p{-}t)\,S^\ell_{(s,\eps),(p-t,-\nu)}
= - t \,S^\ell_{(s,\eps),(t,\nu)} \ .
\ee
With the help of these identities we can rewrite \erf{eq:mod-xfer-S} in
the following form, 
\bea
  \chi_{U^\eps_s}(\tilde q)
  = \sum_{\nu = \pm} S_{(s,\eps),(p,\nu)} \chi_{U^\nu_p}(q) 
  + \sum_{t=1}^{p-1} \big(
  S_{(s,\eps),(t,+)} \psi^+_t(q) +
  S_{(s,\eps),(t,+)}^\ell \varphi^+_t(q) \big)  \ ,
\eear\ee
where $\psi^+_s(q)$ was defined in \erf{eq:psi+-def} and
$\varphi^+_s(q)$ is given by
\be
  \varphi^+_s(q) = - i \tau 
  \frac{2p}{p{-}s} \Big( 
  \frac{p{-}s}{2p} \chi_{U^+_s}(q) 
              - \frac{s}{2p} \chi_{U^-_{p-s}}(q) \Big)
  = - i \tau\, \frac{2p}{p{-}s}\, \frac{\theta'_{p-s,p}(q)}{\eta(q)} 
       \ .
\ee

To determine the boundary states we begin
by considering the overlap of the brane associated to the vacuum
representation $(1,+)$ with itself. Since the fusion of the vacuum
with the vacuum is just the vacuum we have
\bea
  \langle\!\langle (1,+) |\!| 
  q^{\frac{1}{2}(L_0+\bar{L}_0) -  \frac{c}{24}} 
  |\!| (1,+) \rangle\!\rangle 
=  
  \chi_{U_1^+}(\tilde{q}) 
\\ \displaystyle
\qquad =~ 
  \sum_{\nu = \pm} S_{(1,+),(p,\nu)} \chi_{U^\nu_p}(q) 
  + \sum_{t=1}^{p-1} \big(
  S_{(1,+),(t,+)} \psi^+_t(q) +
  S_{(1,+),(t,+)}^\ell \varphi^+_t(q) \big)  \ .
\eear\labl{eq:1+-1+-spec}
The terms from the first sum of the last line come from the
Ishibashi states  
in the sectors $U_p^\pm$, while the contributions with 
$t=1,\ldots, p{-}1$ come from Ishibashi states in $\Hc_{t}$. 
The former Ishibashi states $|U_p^\pm\rangle\!\rangle$ are unique
up to normalisation, and we choose
\be\label{Iship}
\langle\!\langle U_p^\nu | 
q^{\frac{1}{2}(L_0+\bar{L}_0) -  \frac{c}{24}} 
|U_p^\nu \rangle\!\rangle = S_{(1,+),(p,\nu)} \, \chi_{U^\nu_p}(q) 
\ ,
\ee
where the additional factor avoids the introduction of square
roots later on. 
As regards the Ishibashi states coming from $\Hc_t$, 
the analysis in section \ref{sec:Ishi-space} showed that there are
{\it a priori} three independent such Ishibashi states, but it
will turn out that we will need only two. 
Given the validity of \erf{eq:1+-1+-spec} there has to exist
an Ishibashi state $|P_t\rangle\!\rangle$ in $\Hc_t$ such that 
\be
\langle\!\langle P_t | 
q^{\frac{1}{2}(L_0+\bar{L}_0) -  \frac{c}{24}} 
|P_t \rangle\!\rangle 
= S_{(1,+),(t,+)} \, \psi^+_t(q) +
  S_{(1,+),(t,+)}^\ell \, \varphi^+_t(q) \ .
\labl{eq:Ps-Ishi}
Then the boundary state corresponding to the vacuum brane $(1,+)$ is
simply
\be\label{1+}
|\!|(1,+)\rangle\!\rangle = |U_p^+\rangle\!\rangle +  
|U_p^-\rangle\!\rangle 
+ \sum_{t=1}^{p-1} |P_t\rangle\!\rangle 
\ .
\ee
For the general boundary state $|\!|(s,\eps)\rangle\!\rangle$ we now
make the ansatz
\be
|\!|(s,\eps)\rangle\!\rangle = 
B^+_{(s,\eps)}\,|U_p^+\rangle\!\rangle +  
B^-_{(s,\eps)}\,|U_p^-\rangle\!\rangle 
+ \sum_{t=1}^{p-1}\big( B^t_{(s,\eps)} |P_t\rangle\!\rangle 
+ \tilde{B}^t_{(s,\eps)} |U_t\rangle\!\rangle \big)
\ ,
\labl{eq:s-eps-ansatz}
where $|U_t\rangle\!\rangle$ is a second Ishibashi state in $\Hc_t$
whose overlap with $|P_t\rangle\!\rangle$  will be determined shortly. 
By our ansatz \erf{eq:conj-groth} we must have
\bea
  \langle\!\langle (1,+) |\!| 
  q^{\frac{1}{2}(L_0+\bar{L}_0) -  \frac{c}{24}} 
  |\!| (s,\eps) \rangle\!\rangle 
=  
  \chi_{U_s^\eps}(\tilde{q}) 
\\ \displaystyle
 =~ 
  \sum_{\nu = \pm} S_{(s,\eps),(p,\nu)} \chi_{U^\nu_p}(q) 
  + \sum_{t=1}^{p-1} \big(
  S_{(s,\eps),(t,+)} \psi^+_t(q) +
  S_{(s,\eps),(t,+)}^\ell \varphi^+_t(q) \big) \ . 
\eear\labl{eq:1+-s_eps-spec}
This shows that the overlap of $|U_t\rangle\!\rangle$ 
and $|P_t\rangle\!\rangle$ has to be a linear combination of
$\psi^+_t(q)$ and $\varphi^+_t(q)$. For $t=1,\dots,p{-}1$ we have
$S_{(1,+),(t,+)}^\ell \neq 0$ 
so that the second term in \erf{eq:Ps-Ishi} is always
non-zero.\footnote{Note that, on the other hand, 
$S_{(1,+),(p/2,+)}=0$ for $p$ even.}
Thus by redefining 
$|U_t\rangle\!\rangle \mapsto |U_t\rangle\!\rangle 
+ \lambda |P_t\rangle\!\rangle$ 
if necessary, and by rescaling $|U_t\rangle\!\rangle$, we can achieve
that 
\be
\langle\!\langle P_t | 
q^{\frac{1}{2}(L_0+\bar{L}_0) -  \frac{c}{24}} 
|U_t \rangle\!\rangle = \psi^+_t(q) \ .
\ee
Comparing \erf{eq:s-eps-ansatz} and \erf{eq:1+-s_eps-spec} 
now results in the conditions 
$S_{(1,+),(p,\nu)} B^\nu_{(s,\eps)} = S_{(s,\eps),(p,\nu)}$
as well as
\be
  S_{(1,+),(t,+)}^\ell B^t_{(s,\eps)} = S_{(s,\eps),(t,+)}^\ell \ ,
\qquad
S_{(1,+),(t,+)} B^t_{(s,\eps)} + \tilde{B}^t_{(s,\eps)} 
   =S_{(s,\eps),(t,+)} \ .
\ee
The general boundary state is therefore given by
\bea
  |\!|(s,\eps)\rangle\!\rangle
  = \sum_{\nu = \pm} \frac{S_{(s,\eps),(p,\nu)}}{ S_{(1,+),(p,\nu)}}
  |U^\nu_p\rangle\!\rangle
  + \sum_{t=1}^{p-1} 
  \frac{S^\ell_{(s,\eps),(t,+)}}{ S^\ell_{(1,+),(t,+)}} |P_t\rangle\!\rangle
  \enl
  \hspace*{5em} + \sum_{t=1}^{p-1} \Bigg(
    S_{(s,\eps),(t,+)} - S_{(1,+),(t,+)}  
    \frac{S^\ell_{(s,\eps),(t,+)}}{ S^\ell_{(1,+),(t,+)} } 
   \Bigg) |U_t\rangle\!\rangle  \ .
\eear\labl{eq:bnd-state-s_eps}
Before we can verify that this indeed reproduces \erf{eq:conj-groth}
we have to compute the self-overlap of $|U_t\rangle\!\rangle$. Let us
set $\langle\!\langle U_t | 
q^{\frac{1}{2}(L_0+\bar{L}_0) -  \frac{c}{24}} 
|U_t \rangle\!\rangle = h_t(q)$. Then from \erf{eq:bnd-state-s_eps}  
and \erf{eq:mod-S-mat} it follows that
\bea
  \langle\!\langle (p,+) |\!| 
  q^{\frac{1}{2}(L_0+\bar{L}_0) -  \frac{c}{24}} 
  |\!| (p,+) \rangle\!\rangle =
  \sqrt{2/p} \, \big( \chi_{U^+_p}(q) - (-1)^p \chi_{U^-_p}(q) \big)
  + \sum_{t=1}^{p-1} \tfrac{2}{p}h_t(q)  
  \enl  
  \quad =~ \sum_{t=1}^{p-1} \tfrac{2}{p}h_t(q) + 
  \begin{cases}
  2\sum_{r=1;2}^{p-1} \big( \chi_{U^+_r}(\tilde q)+
    \chi_{U^-_{p-r}}(\tilde q) \big) & \text{if}~p~\text{even}~,
  \\
  \chi_{U^+_p}(\tilde q)+
  2\sum_{r=1;2}^{p-2} \big( \chi_{U^+_r}(\tilde q)+
    \chi_{U^-_{p-r}}(\tilde q) \big) & \text{if}~p~\text{odd}~,
  \end{cases}
\eear\labl{eq:p-p-spec}
where the `;2' in the sum means that the sum is taken in steps of
two.
This should be equal to \erf{eq:conj-groth} for 
$(s_1,\eps_1)=(s_2,\eps_2)=(p,+)$. Note that
the fusion of $U_p^+$ with itself is 
\be
U_p^+ \circledast U_p^+ = \left\{ 
\begin{array}{ll}
{\displaystyle 
P_1^{+} \oplus P_3^{+} \oplus \cdots \oplus P_{p-1}^{+}} \qquad
& \hbox{if $p$ even} \\[4pt]
{\displaystyle 
P_1^{+} \oplus P_3^{+} \oplus \cdots \oplus P_{p-2}^{+} \oplus
U_p^+ } \qquad
& \hbox{if $p$ odd} 
\end{array}
\right.
\ee
(see appendix~\ref{app:fusion}). Comparing the character of this
and \erf{eq:p-p-spec} fixes $h_t(q)=0$, so that
\be
\langle\!\langle U_t | 
q^{\frac{1}{2}(L_0+\bar{L}_0) -  \frac{c}{24}} 
|U_t \rangle\!\rangle = 0 \ .
\ee
Now that the overlaps of the Ishibashi states and the boundary states
are fixed we can perform the consistency check of our ansatz by
substituting \erf{eq:bnd-state-s_eps} into \erf{eq:conj-groth}.
This results in the expression
\bea
  N_{(s_1,\eps_1)(s_2,\eps_2)}^{\hspace{3em}(r,\mu)} 
  = \sum_{\nu=\pm}
  \frac{S_{(s_1,\eps_1),(p,\nu)}S_{(s_2,\eps_2),(p,\nu)}
     S_{(p,\nu),(r,\mu)}}{S_{(1,+),(p,\nu)}}
  \enl \hspace*{7em}
  + \sum_{t=1}^{p-1} \sum_{\nu=\pm}
  \Bigg(
  \frac{S_{(s_1,\eps_1),(t,\nu)}
  S^\ell_{(s_2,\eps_2),(t,\nu)}
  +
  S^\ell_{(s_1,\eps_1),(t,\nu)}
  S_{(s_2,\eps_2),(t,\nu)}
  }{
  S^\ell_{(1,+),(t,\nu)}}
  \enl \hspace*{13em}
  - 
  \frac{
  S^\ell_{(s_1,\eps_1),(t,\nu)}
  S^\ell_{(s_2,\eps_2),(t,\nu)}
  (i\tau + S_{(1,+),(t,\nu)}/S^\ell_{(1,+),(t,\nu)})
  }{
  S^\ell_{(1,+),(t,\nu)}}
  \Bigg)
  \enl \hspace*{17em}
    \times
  \Big( S_{(t,\nu),(r,\mu)} 
  - i \big(\tfrac{-1}{\tau}\big) S^\ell_{(t,\nu),(r,\mu)} \Big)
  \ .
\eear\labl{eq:verlinde-1}
In particular, the $\tau$-dependence on the right hand side
has to cancel.
We have verified numerically in a large number of examples that
the right hand side of \erf{eq:verlinde-1} 
indeed reproduces the structure constants of the
Grothendieck ring as determined by \erf{eq:Wp-Groth-ring}.

While equation \erf{eq:verlinde-1} still looks quite complicated,
it can be simplified considerably in the following way. 
For $t=1,\dots,p{-}1$ consider the 
$S_{(s,\eps),(t,\nu)}$ and $S^\ell_{(s,\eps),(t,\nu)}$ as formal 
variables and introduce a derivation $D$ on these by setting
$D[S^\ell]=S$ and $D[S]=0$, {\it i.e.}
\be
  D\big[ f(S^\ell_{(s,\eps),(t,\nu)}) \big] 
  = f'(S^\ell_{(s,\eps),(t,\nu)}) S_{(s,\eps),(t,\nu)} \ , \qquad
  D\big[ f(S_{(s,\eps),(t,\nu)}) \big] = 0 \ .
\ee
Then one can {\it e.g.}\ write \erf{eq:mod-xfer-S} as
\be
  \chi_{U^\eps_s}(\tilde q)
  = \sum_{\nu = \pm} S_{(s,\eps),(p,\nu)} \chi_{U^\nu_p}(q) +
   \sum_{t=1}^{p-1} \sum_{\nu = \pm} \big( D - i \tau \, \id \big)
  \big[ S^\ell_{(s,\eps),(t,\nu)} \big] \cdot
  \chi_{U^\nu_t}(q) ~~.
\labl{eq:character-D}
Equation \erf{eq:verlinde-1} can then be written more compactly
as
\bea
  N_{(s_1,\eps_1)(s_2,\eps_2)}^{\hspace{3em}(r,\mu)} 
  = \sum_{\nu=\pm}
  \frac{S_{(s_1,\eps_1),(p,\nu)}S_{(s_2,\eps_2),(p,\nu)}
     S_{(p,\nu),(r,\mu)}}{S_{(1,+),(p,\nu)}}
  \enl \hspace*{2em}
  + \sum_{t=1}^{p-1} \sum_{\nu=\pm}
  \big( D - i \tau \, \id \big)
  \Bigg[
  \frac{S^\ell_{(s_1,\eps_1),(t,\nu)}S^\ell_{(s_2,\eps_2),(t,\nu)}}{
   S^\ell_{(1,+),(t,\nu)}}\Bigg] \cdot
  \big( D - i \big(\tfrac{-1}{\tau}\big) \id \big)
  \Big[ S^\ell_{(t,\nu),(r,\mu)} \Big]  ~~,
\eear\labl{eq:verlinde-2}
where the $\tau$-dependent terms on the right hand side cancel.
This formula can be understood as a logarithmic version of the
Verlinde formula \cite{Verlinde:1988sn},
which describes the structure constants of
the Grothendieck ring in non-logarithmic
rational conformal field theories (of course, for these the
Grothendieck ring coincides with the fusion rules of the irreducible
representations).  
Verlinde-like formulas for $\Wc_p$-representations have also
been studied in \cite{Flohr95,Fuchs:2003yu,Flohr:2007jy}.

As opposed to the procedure in \cite{Fuchs:2003yu} 
resting on block-diagonalising
the fusion rules,\footnote{ 
The matrix $\mathsf{S}$ appearing in the 
block-diagonalisation is related to $S$ and $S^\ell$ via
$\mathsf{S} = S + S^\ell$
(see \cite[eqns.\,(3.3), (4.12) and (5.17)]{Fuchs:2003yu}).}
the formula \erf{eq:verlinde-2} does not involve any
choices; the matrices $S$ and $S^\ell$ are uniquely fixed by 
\erf{eq:mod-xfer-S}.
Furthermore, in the given form it is quite suggestive
how to generalise \erf{eq:verlinde-2}: the sum runs over all
irreducible representations and each summand in  \erf{eq:verlinde-2}
looks as in the usual Verlinde formula, 
but with additional $D$-operators inserted, where
the number of insertions is related to the size of the Jordan cell of
$L_0$ in the corresponding projective cover.
(For the $\Wc_p$-representations these Jordan cells are all of
length one or two.)

\sect{Conclusion and Outlook}\label{sec:concl}

In this paper we have proposed a simple method to compute the space of
bulk fields for a logarithmic rational conformal field theory. The
construction starts from the assumption that there is a boundary
condition whose space of boundary fields consists only of the chiral
algebra $\Vc$ itself. 
The space of bulk fields is then the largest space that can be
coupled to the space of boundary fields in a non-degenerate way,
consistent with the action of $\Vc$. When applied to non-logarithmic
rational conformal field theories, this construction yields the 
charge-conjugation modular invariant theory.   

We verified that our method gives a modular invariant partition
function when applied to the $c_{1,p}$ triplet models. As a
consistency check of the ansatz we computed the set of boundary states
--- one for each irreducible representation of the $\Wc_p$-algebra ---
and checked that their overlaps give consistent amplitudes in the open
channel.
We also confirmed that there is indeed a boundary condition whose
space of boundary fields is given by $\Vc$. 
The analysis of the boundary states finally led to a Verlinde-like formula
for the structure constants of the Grothendieck ring of the
$\Wc_p$-representations. We also conjectured a formula for the fusion
rules of these representations.

\medskip

\noindent There are a number of questions that deserve further study: 
\begin{list}{-}{\topsep .4em \itemsep 0em \leftmargin 1em}
\item
While we were able to show that the partition function
is invariant under the $T$-modular transformation,
$Z(\tau) = Z(\tau{+}1)$, it remains to
prove in general that it is also invariant under
the $S$-transformation, $Z(\tau) = Z(-1/\tau)$.  
\item
The ansatz for the spaces of boundary fields given in \erf{conj} does determine
the character of the corresponding $\Wc_p$-representation, but not the
representation itself. On the other hand, the tensor product
\erf{eq:irrep-fusion} of irreducible representations provides a
natural conjecture for the $\Wc_p$ action on the spaces of boundary
fields. For $p=2$ this has been verified to some extent in
\cite{Gaberdiel:2006pp}, and it would be good to check that this
remains true for all $p$. 
\item
To have a consistent conformal field theory one also has to find a set
of structure constants that satisfy the sewing constraints. For
non-logarithmic rational theories such a set of structure constants is
uniquely determined by the boundary theory  
\cite{Fuchs:2004xi,Fjelstad:2006aw}, and it would be interesting to
understand to which 
extent this remains true for logarithmic models.
\item
As already noted in \cite[sect.\,6.2]{Quella:2007hr}, the result
\erf{eq:Wp-bulk-space} for the space of bulk states bears a remarkable
resemblance to the decomposition of the regular representation of 
$\bar{\mathcal{U}}_q s\ell(2)$, 
see \cite[prop.\,4.4.2]{Feigin:2005zx}. Understanding this relation
better might help to formulate the construction presented in this
paper on a purely categorical level without explicit mention of the
action of $\Vc$. 
\item The analysis of WZW models with supergroup targets in
\cite{Schomerus:2005bf,Gotz:2006qp,Saleur:2006tf,Quella:2007hr} uses 
quite a different starting point to obtain the bulk space
as compared to our construction, namely
harmonic analysis on supergroups. It would be interesting to evaluate
our quotient expression for $\Hc_\text{bulk}$ 
(\ref{eq:bulk-general}) 
for these supergroup models, and see if the result agrees with their
findings. 
\item 
It would be good to understand the precise relation between
\cite{Flohr95,Fuchs:2003yu,Flohr:2007jy} and 
our formula \erf{eq:verlinde-2},
in particular since we recover the structure constants of the
Grothendieck ring determined in \cite{Fuchs:2003yu}.
Also, it would be very interesting to see if 
\erf{eq:verlinde-2}, with the modifications suggested
there, determines the Grothendieck ring of other logarithmic conformal
field theories as well.
\end{list}
We hope to return to some of these points in the near
future.

\subsection*{Acknowledgements}

We thank J\"urgen Fuchs, Thomas Quella and Volker Schomerus for useful
discussions and comments on the draft,
and Michael Flohr and Holger Knuth for helpful
correspondences. 
IR thanks the ETH Z\"urich and the University of Hamburg 
for hospitality, where part
of this work was carried out.
The research of MRG was partially supported by the
Swiss National Science Foundation and the Marie Curie network
`Constituents, Fundamental Forces and Symmetries of the Universe'
(MRTN-CT-2004-005104). The research of IR was partially supported by
the EPSRC First Grant EP/E005047/1, the PPARC rolling grant
PP/C507145/1 and the Marie Curie network `Superstring Theory'
(MRTN-CT-2004-512194).

\appendix

\sect{Compatibility of $b$ with the
  $\Vc$-action}\label{app:b-mode-com} 

In this appendix we describe the invariance condition that defines  
the space 
$B(\hat\Hc)$ for a given $\VxV$-representation $\hat\Hc$. This
condition comes directly from the definition of conformal blocks as
co-invariants with respect to an 
action of the conformal vertex algebra, see 
\cite{Frenkel-BenZvi-2nd-Edition}.

Let $W$ be a quasi-primary element of $\Vc$ of conformal weight $h_W$.
Let $\mathcal{F}(h_W)$ be the space of functions that are meromorphic
on $\Cb$, holomorphic on $\Cb - \{i,0,-i\}$, and behave as
$(\text{const}) z^{2h_W-2}$ at infinity. Given an element
$f \in \mathcal{F}(h_W)$ and a point $x \in \Cb$ we define the formal
sum of modes of $W$
\be
   W[f,x] = \sum_{m \in \Zb} a_{m+h_W-1} W_{m} \ ,
   \qquad \text{where}
   \quad
   f(x{+}\zeta) = \sum_{m \in \Zb} a_m \zeta^m \ .
\labl{eq:W[f,x]-def}
With this definition $W[z^{m+h_W-1},0] = W_m$, and
$z^{m+h_W-1} \In \mathcal{F}(h_W)$ for $m \le h_W{-}1$, {\it i.e.}\  
precisely when $\langle 0 |W_m = 0$. For the same reason, 
$\langle 0 |W[f,0] = 0$ for all $f \in \mathcal{F}(h_W)$. The
invariance condition for the conformal three-point blocks $\beta$ with
insertions of $M$, $N$, $\Vc$ at $i,- i,0$, respectively, is then
obtained by inserting the contour integral $\oint f(z) W(z) dz$ around
infinity and  deforming the contour. One obtains
\be
   \beta\big(\, W[f,i]p,\,q,\,v \,\big)
   + \beta\big(\, p,\,W[f,-i]q,\,v \,\big)
   + \beta\big(\, p,\,q,\,W[f,0]v \,\big) ~=~ 0 \ ,
\ee
where $W \In \Vc$ is quasiprimary, and $f \In \mathcal{F}(h_W)$,  
$p \In M$, $q \In N$ and $v \In \Vc$ are arbitrary.

This translates into the following definition for $B(\hat \Hc)$:
given a $\VxV$-representation $\hat \Hc$, the
space $B(\hat \Hc)$ consists of all bilinear maps
$b : \hat \Hc \ti \Vc \rightarrow \Cb$ with the property that for all
$u \In \hat\Hc$, $v \In \Vc$, and for all quasi-primary $W \In \Vc$ and
all $f \In \mathcal{F}(h_W)$,
\be
   b\big(\, W[f,i] u,\,v \,\big)
   + b\big(\, \bar W[f,-i] u,\,v \,\big)
   + b\big(\, u,\,W[f,0]v \,\big) ~=~ 0 \ .
\labl{eq:bilin-Uhat-V-cond}
Here, by $\bar W[f,x]$ we mean the formal sum
$\sum_{m \in \Zb} a_{m+h_W-1} \bar W_{m}$, where the coefficients 
$a_m$ are defined as in \erf{eq:W[f,x]-def}.

\sect{The kernel of $\bb$ is a $\VxV$-representation} 
\label{app:kernel-of-b}

Given a $\VxV$-representation $\hat U$ together with a pairing 
$b \in B(\hat U)$ we will prove that the kernel 
of $\bb : \hat U \rightarrow \Vc^*$ is a  
$\VxV$-subrepresentation of $\hat U$. This statement is implied by the
following lemma. 

\medskip
\noindent
{\bf Lemma.} Let $W \In \Vc$ be quasi-primary and let $u \in \hat U$ be 
such that $b(u,v)=0$ for all $v \In \Vc$. Then also 
$b(W_m u,v) = 0 = b(\bar W_m u,v)$ for all $v \In \Vc$ and $m \In \Zb$.

\medskip
\noindent
{\bf Proof.} We will prove the assertion by induction on the mode 
number $m$. By definition of a $\VxV$-representation, for every vector
$u$ there is an integer $M(u)$ such that $W_m u = 0 = \bar W_m u$ for
all $m \ge M(u)$. To start the induction we note that
$b(W_m u,v) = 0 = b(\bar W_m u,v)$ for all $v \In \Vc$ and $m \ge M(u)$.
Suppose now we have proved the statement for all $m > m_0$. Consider
the function 
\be
f(z) = (z{-}i)^{m_0+h_W-1} ((z{+}i)/(2i))^{M(u)+h_W-1} 
(z/i)^{-m_0-M(u)} ~ \in \mathcal{F}(h_W) ~.
\ee
Then $W[f,i] = W_{m_0} + (\text{higher})$, where $(\text{higher})$
stands for terms with mode number greater than $m_0$. Furthermore we
have $W[f,-i] = (\text{const})\cdot W_{M(u)} + (\text{higher})$  
so that $\bar W[f,-i] u = 0$.
We compute
\be\begin{array}{ll}
  b(W_{m_0} u , v) \etb\!\!
  \overset{(1)}{=}~ b( W[f,i] u , v) - b( (\text{higher}) u , v)
  ~\overset{(2)}{=}~ b( W[f,i] u , v) 
  \\[.2em]
  \etb\!\!
  \overset{(3)}{=}~ - b( \bar W[f,-i] u , v) - b( u , W[f,0] v) 
  ~\overset{(4)}{=}~ 0 ~.
\eear\ee
In step 2 we employed the induction assumption to set the second term
to zero, step 3 uses the definition of $B(\hat U)$, and in step 4 the
first term vanishes because $\bar W[f,-i] u = 0$ and the second term  
is of the form $b(u,v')$ for some $v' \In \Vc$, which is zero by
assumption. Similarly one can show that $b(\bar W_{m_0} u , v) = 0$ 
for all $v$. \hfill $\Box$

\sect{Fusion rules for
  $\mathcal{W}_p$-representations}\label{app:fusion} 

The product of irreducible representations
in the Grothendieck ring has been conjectured in \cite{Fuchs:2003yu}.
In this appendix we extend this conjecture to 
the fusion product $\circledast$ of the irreducible and
projective representations.

\medskip

The (additive) Grothendieck group $K_0(\Cc)$ 
of an abelian category $\Cc$ is the quotient
of the free abelian group generated by isomorphism classes $[U]$ of
objects in $\Cc$ by the subgroup generated by
the relations $[K] + [Q] = [M]$ for each short exact sequence 
$0 \rightarrow K \rightarrow M \rightarrow Q \rightarrow 0$. 
If $\Cc$ is also monoidal and the tensor 
(fusion) functor $\circledast$ 
is exact, then we obtain a ring structure on $K_0(\Cc)$ via
$[U] \cdot [V] = [U \circledast V]$.

Denote the category of $\Wc_p$-modules 
by $\Cc_p$. The Grothendieck group
is freely generated by the $2p$ classes of the irreducible representations
$[U^\pm_s]$, $s=1,\dots,p$. From \cite[sect.\,2.4]{Fuchs:2003yu} we
have the exact sequences 
\be
 0 \rightarrow U^\nu_s \rightarrow M^\nu_{\eps,s} 
   \rightarrow U^{-\nu}_{p-s} \rightarrow 0
 \quad , \quad
 0 \rightarrow U^\nu_s \rightarrow N^\nu_{s} 
   \rightarrow U^{-\nu}_{p-s} \oplus U^{-\nu}_{p-s} \rightarrow 0 \ ,
\ee
where $s=1,\dots,p{-}1$ and $\nu,\eps = \pm$; in addition we have 
the first sequence of \erf{eq:exact-P-middle}. These give
the following identities in $K_0(\Cc_p)$,
\be
  [P^\nu_s] = 2[U^\nu_s]+2[U^{-\nu}_{p-s}] \ ,\qquad
  [N^\nu_s] = [U^\nu_s]+2[U^{-\nu}_{p-s}] \ ,\qquad
  [M^\nu_{\pm,s}] = [U^\nu_s]+[U^{-\nu}_{p-s}] \ .
\labl{eq:K0-identities}
Since the exact sequences split when considered as sequences of
graded vector spaces (and not as sequences of $\Wc_p$-modules) the
identities \erf{eq:K0-identities} are sum rules for the characters
of the corresponding representations.

Since $M^\nu_{\pm,s}$ are submodules of $N^\nu_s$, the exact
sequence $0 \rightarrow M^\nu_{\pm,s} \rightarrow N^\nu_s 
   \rightarrow X \rightarrow 0$ implies the relation
$[N^\nu_s] = [M^\nu_{\pm,s}] + [X]$, which together with 
\erf{eq:K0-identities} shows $X \cong U^{-\nu}_{p-s}$.
A similar argument shows that 
in $0 \rightarrow M^\nu_{\eps,s} \rightarrow P^\nu_s 
   \rightarrow Y \rightarrow 0$ we either have
$Y \cong M^{-\nu}_{\pm\eps,p-s}$ 
(since the quotient $N^\nu_s/M^\nu_{\eps,s} \cong U^{-\nu}_{p-s}$ is
embedded in $P^\nu_s/M^\nu_{\eps,s} \cong Y$)
or $Y \cong U^\nu_s \oplus U^{-\nu}_{p-s}$.
The second possibility is excluded since $P^\nu_s$ is already the
projective cover of $U^\nu_s$. The choice of sign in 
$Y \cong M^{-\nu}_{\pm\eps,p-s}$ is a convention which can be reversed
by redefining 
$M^{\pm,\text{new}}_{\eps,s} = M^{\pm,\text{old}}_{\pm\eps,s}$.
We fix the convention as stated in \erf{eq:exact-P-middle}.

\medskip

Assuming that the tensor (fusion) functor on $\Cc_p$ is exact,
in \cite{Fuchs:2003yu} the following conjecture
for the ring structure on $K_0(\Cc_p)$ is made (we follow the
exposition in \cite[sect.\,6.3]{Fuchs:2006nx}). The product is
commutative, and ordering the factors such that $1 \le t \le s \le p$
we have 
\be
  [U_s^\mu] \cdot [U_t^\nu] = \sum_{i=1}^t [\hat U^{\mu\nu}_{s-t+2i-1}]
  \ ,\qquad
  [\hat U^\pm_x] =
  \begin{cases}
  [U^\pm_x] & \text{for} ~ 1 \le x \le p  ~,\\
  [U^\pm_{2p-x}]+2[U^\mp_{x-p}] & \text{for} ~ 
	p{+}1 \le x \le 2p{-}1 \ .
  \end{cases}
\labl{eq:Wp-Groth-ring}

The Grothendieck ring does not determine the fusion product of
representations uniquely. However, we can arrive at a convincing
proposal for the fusion product of irreducible and projective
representations using the the analysis of the fusion of Virasoro
(rather than $\Wc_p$) representations in \cite{GabKau96a}. This leads
to the following natural ansatz for the fusion product of two
irreducible $\Wc_p$-representations, 
\be
  U^{\mu}_s \circledast U^{\nu}_t
  ~= 
  \hspace{-2em}\bigoplus_{r=|s{-}t|+1;2}^{\min(s+t-1,2p-s-t-1)}
  \hspace{-2em} U^{\mu\nu}_r
  ~\oplus
   \hspace{-1em} \bigoplus_{r=2p-s-t+1;2}^{M} \hspace{-1em} 
         P^{\mu \nu}_r
  \qquad \text{where} \quad
  M = \begin{cases} p{-}1 & \text{if}~p{+}s{+}t ~\text{even} \ , \\
                    p & \text{if}~p{+}s{+}t ~\text{odd} \ . 
      \end{cases}
\labl{eq:irrep-fusion}
The `$;2$' means the above direct sums are taken in steps of 2.
On the level of the Grothendieck ring \erf{eq:irrep-fusion} is
equivalent to \erf{eq:Wp-Groth-ring}.
Note also that $U^-_1$ is a simple current, 
$U^-_1 \circledast U^\eps_s = U^{-\eps}_s$.

According to proposition 2.2 in \cite{Etingof2003}, 
tensor (fusion) products involving at least one projective module 
are already fixed by the Grothendieck ring
(to apply this result we need
to assume that $\Cc_p$ is a finite tensor category, see
\cite{Etingof2003} for details). The proposition states
that
\be
  P_i \circledast Z =  \bigoplus_{j,k \in \Irr} N^{~i}_{kj} 
   \,[Z : U_j]\, P_k \ ,
\labl{eq:other-fusion}
where, for $j\in\Irr$, $U_j$ is the simple object with label $j$,
$P_j$ its projective cover, and $Z$ an arbitrary object in $\Cc_p$.
The $N_{kj}^{~i}$ are the structure constants of the Grothendieck
ring \erf{eq:Wp-Groth-ring}, 
$[U_k] \cdot [U_j] = \sum_{i \in \Irr} N_{kj}^{~i} [U_i]$ 
and $[Z:U_j] \in \Zb_{\ge 0}$  gives the decomposition of $Z$ 
in $K_0(\Cc)$ as $[Z] = \sum_{j\in\Irr} [Z:U_j] \, [U_j]$.

In writing \erf{eq:other-fusion} we have assumed that the simple objects
are self-dual, {\it i.e.}\ that for the 
$\Wc_p$-representations we have
$U^{\pm\,*}_s \cong U^\pm_s$. (The statement without this assumption
can be found in \cite[prop.\,2.2]{Etingof2003}.)
We will also assume that $P^{\pm\,*}_s \cong P^\pm_s$.

Equations \erf{eq:irrep-fusion} and \erf{eq:other-fusion} determine
now all fusion products 
$U^\mu_s \circledast U^\nu_t$,
$U^\mu_s \circledast P^\nu_t$, and
$P^\mu_s \circledast P^\nu_t$ uniquely.
Explicitly, we find that
\be
  U^{\mu}_s \circledast P^{\nu}_t
  ~= \hspace*{-1.5em}
\bigoplus_{r=|s{-}t|+1;2}^{\min(s+t-1,2p-s-t-1)}
   \hspace*{-2.5em}P^{\mu\nu}_r
  ~~\oplus \hspace*{-.5em}
    \bigoplus_{r=2p-s-t+1;2}^{M} \hspace{-1em} 
         \!\! 2 P^{\mu \nu}_r 
  ~\oplus \hspace*{-.5em}
    \bigoplus_{r=p+1+t-s;2}^{\hat{M}} \hspace*{-1em} 2 P^{-\mu\nu}_r
  \quad \text{for}  ~~t\leq p{-}1\ ,
\labl{eq:fusion-expl}
where $M$ is defined as in (\ref{eq:irrep-fusion}) and
\be
\hat{M} = \left\{
\begin{array}{ll}
p{-}1 \quad & \hbox{if $s+t$ even} \\
p \qquad & \hbox{if $s+t$ odd.} 
\end{array}
\right.
\ee
Finally,   
\be
P_s^{\mu} \circledast P_t^{\nu}  
~= 2 \, U_s^{\mu} \circledast P_t^{\nu} \oplus
2 \, U_{p-s}^{-\mu} \circledast P_t^{\nu} 
\qquad \text{for}~~
s,t\leq p{-}1\ ,
\labl{eq:fusion2}
where the right-hand-side is defined by (\ref{eq:fusion-expl}).
It is straightforward to check that the fusion product defined by 
\erf{eq:irrep-fusion}, \erf{eq:fusion-expl} and \erf{eq:fusion2}
is compatible with the product \erf{eq:Wp-Groth-ring}
of the Grothendieck ring. We have also tested for a large number of
values for $p$ that the fusion product is associative (as it must be).
\medskip

Let us also compare our proposal for the fusion products with the
results in \cite[sect.\,6]{Flohr:2007jy}, which are based on
generalised versions of the Verlinde formula. Formula
\erf{eq:irrep-fusion} agrees with \cite[eqn.\,(6.16)]{Flohr:2007jy}
if we identify $U^+_s = [h_{1,s}]$, $U^-_s = [h_{1,3p-s}]$
and $P^+_s = [\widetilde h_{1,2p-s}]$. 
However, the method used in \cite{Flohr:2007jy} does not distinguish 
between $P^+_s$ and $P^-_{p-s}$. Keeping this in mind, one can recover
\cite[eqn.\,(6.19)]{Flohr:2007jy} by starting from
\erf{eq:fusion-expl} and replacing in addition
$P^-_s \mapsto [\widetilde h_{1,p+s}]$.

\medskip

To find the dimension of the intertwiner spaces
$\Hom_\Vc(P^\mu_s,P^\nu_t)$ one can proceed as follows. First note
that by the properties of duals and by uniqueness of the projective
cover we have 
\be
\Hom_\Vc(P^\mu_s,P^\nu_t) \cong \Hom_\Vc(P^\mu_s \circledast 
P^{\nu\,*}_t, U^+_0)
\ , \qquad
  \dim\Hom_\Vc(P^\mu_s,U^+_0) = \delta_{\mu,+} \, \delta_{s,0} \ .
\ee
As mentioned above we assume that $P^{\pm\,*}_s \cong P^\pm_s$.
To obtain the dimension of $\Hom_\Vc(P^\mu_s,P^\nu_t)$ it is thus
sufficient to compute the multiplicity of $P^+_0$ in 
$P^\mu_s \circledast P^\nu_t$. 
From \erf{eq:other-fusion} we find
\be
 P^\mu_s \circledast P^\nu_t ~\cong~ [P^\nu_t:U^\mu_s] \,P^+_0 ~\oplus~
  \text{(other projectives)}\ .
\ee
The multiplicity in $K_0(\Cc_p)$ follows from \erf{eq:K0-identities}
to be
$[P^\nu_t:U^\mu_s] = 2\delta_{\nu,\mu} \delta_{s,t}
+2\delta_{\nu,-\mu} \delta_{s,p-t}$. This shows that the dimension of
the intertwiner spaces is indeed as proposed in \erf{interitw}.

\sect{The structure of $\hat\Hc_s / \Nc_s$}\label{app:proj-inj-ker} 

In this appendix we want to prove \erf{decomp1}. To do so we 
recall that $N^\eps_s$ is a subrepresentation of
$P^\eps_s$. In each generalised eigenspace $(P^\eps_s)_h$ of
$P^\eps_s$ of eigenvalue $h$ choose a sub-vector space $(V^\eps_s)_h$
such that $(P^\eps_s)_h = (V^\eps_s)_h \oplus (N^\eps_s)_h$. In words,
$(V^\eps_s)_h$ and $(N^\eps_s)_h$ have intersection $\{0\}$ and
together span $(P^\eps_s)_h$. We will write 
$V^\eps_s = \bigoplus_{h   \in \Rb} (V^\eps_s)_h$. 
The vector space $V^\eps_s$ is not a 
$\Wc_p$-subrepresentation of $P^\eps_s$.

Consider the projectors 
$\Pi^\eps_s : P^\eps_s \rightarrow P^\eps_s$ which act as the identity
on $N^\eps_s$, and as zero on $V^\eps_s$ (these are not intertwiners
of the $\Wc_p$-action). Using the $\Pi^\eps_s$ we can define a
projector $\Pi_s : \hat\Hc_s \rightarrow \hat\Hc_s$ by 
$\Pi_s = (\Pi^+_s \otimes \bar\id) \oplus 
(\Pi^-_{p-s} \otimes \bar\id)$. By construction we have 
\be\label{D1}
  \ker(\Pi_s) = (V_s^+ \oti \bar P^{+\,*}_s) \oplus 
  (V^-_{p-s} \oti \bar P^{-\,*}_{p-s}) \ .
\ee
It is proved in the following subsection that the restriction
of $\Pi_s$ to $\Nc_s$ is injective. This implies that 
$\Nc_s \cap \ker(\Pi_s) = \{ 0 \}$.  
Consider the quotient $\hat\Hc_s/\Nc_s$, and for an element 
$x \in \hat\Hc_s$ denote the class in $\hat\Hc_s/\Nc_s$ by $[x]$. We
will show that 
every element of $\hat\Hc_s/\Nc_s$ can be written as $[k]$ with
$k \in \ker(\Pi_s)$. This then implies that $\Nc_s$ and
$\ker(\Pi_s)$ together span $\hat\Hc_s$.   

It is enough to consider elements of $\hat\Hc_s$ of the form 
$(v{+}\eta)\otimes \bar q$ where either 
$v \in V^+_s$, $\eta \in N^+_s$,  
$\bar q \in \bar P^{+\,*}_s$, or 
$v \in V^-_{p-s}$, $\eta \in N^-_{p-s}$,  
$\bar q \in \bar P^{-\,*}_{p-s}$. Take the first case, for
concreteness.  
The map $e_\nu : P^-_{p-s} \rightarrow P^{+}_{s}$ has image
$M^{+}_{\nu,s}$ and kernel $M^{-}_{\nu,p-s}$. This implies that
$e_\nu$ maps $N^-_{p-s} \subset P^-_{p-s}$ to $U^{+}_{s}$ and that we
can write an arbitrary element 
$m_\nu \in M^{+}_{\nu,s} \subset P^+_s$ as $m_\nu = e_\nu(w) + u$ for
appropriate $w \in V^-_{p-s}$, $u \in U^{+}_{s}$. The element $u$ in
turn can be expressed as $u = e_-(e_+(w'))$ for some 
$w' \in V^{+}_{s}$. Altogether we see that
for any $\eta \in N^+_s$
\be
  \exists~~
  w_+,w_- \In  V^-_{p-s}~,~w_0 \In V^+_s ~~:~~
  \eta = e_+(w_+) + e_-(w_-) + e_-(e_+(w_0)) \ .
\ee
Since the images of $\id \oti \bar e_\nu^* - e_\nu \oti \bar \id$ are
in $\Nc_s$, it follows that 
in the quotient space $\hat\Hc_s / \Nc_s$  
$[\eta \oti \bar q] = 
[w_+ \oti \bar e_+^*(\bar q)] +
[w_- \oti \bar e_-^*(\bar q)] +
[w_0 \oti \bar e_+^*(( \bar e_-^*(\bar q))]$.
Thus $[(v{+}\eta)\otimes \bar q]$ can be written as a sum of four
terms all of which lie in $\ker(\Pi_s)$. 
Thus we have shown that 
\be
  \hat\Hc_s = (V_s^+ \oti \bar P^{+\,*}_s )\oplus 
  (V^-_{p-s} \oti \bar P^{-\,*}_{p-s}) \oplus \Nc_s
\labl{eq:Hs-decomp-aux1}   
as a vector 
space with generalised $(L_0,\bar L_0)$-grading. Since
$\Pi_s$ commutes with the action of $\bar \Wc_p$ and since
$\Nc_s$ is a 
$\Wc_p{\times}\bar\Wc_p$-subrepresentation, the
decomposition \erf{eq:Hs-decomp-aux1} is also preserved
by the $\bar \Wc_p$-action (but not by the $\Wc_p$-action). 

Finally, we need to show that $V^\eps_s$
is isomorphic (as a graded vector space) to $U^\eps_s$. 
To see this we
observe that the surjection 
$\pi_s^\eps : P^\eps_s \twoheadrightarrow U^\eps_s$ in
the first exact sequence in \erf{eq:exact-P-middle} has kernel
$N^\eps_s$. In particular, $\pi_s^\eps$ restricts to a bijection
$V^\eps_s \rightarrow U^\eps_s$ which is compatible with the
generalised $L_0$-grading (but not with the action of the
$\Wc_p$-modes, or even with the action of $L_0$ itself). This 
then proves \erf{decomp1}.  

\subsection{The projection $\Pi_s$ is injective
on $\Nc_s$}

Recall the decomposition 
$(P^\eps_s)_h = (V^\eps_s)_h \oplus (N^\eps_s)_h$
chosen above, and the inclusions
\erf{eq:rep-include}. Let us also choose subspaces
$(U^\eps_{s})_h$ and $(S^\eps_{\pm,s})_h$ of $(N^\eps_s)_h$
such that $(U^\eps_{s})_h$ is the generalised $L_0$-eigenspace of 
eigenvalue $h$ of $U_s^\eps \subset P_s^\eps$, as well as
$(M^\eps_{\pm,s})_h = (S^\eps_{\pm,s})_h \oplus (U^\eps_{s})_h$.
As was the case for $V_s^\eps$, the subspaces 
$S^\eps_{\pm, s} = \bigoplus_{h \in \Rb} (S^\eps_{\pm,s})_h$
are not $\Wc_p$-submodules. We have now
chosen the direct sum decompositions
\be
  P_s^\eps = V^\eps_{s} \oplus S^\eps_{+,s} \oplus 
    S^\eps_{-,s} \oplus U^\eps_{s} \ , \qquad
  N_s^\eps = S^\eps_{+,s} \oplus 
    S^\eps_{-,s} \oplus U^\eps_{s} \ , \qquad
  M_{\pm,s}^\eps = S^\eps_{\pm,s} \oplus U^\eps_{s} \ .
\labl{eq:P-direct-sum}

According to the construction in section \ref{sec:bulk-construct}
every element $k$ of $\Nc_s$ can be written as a sum of the form
\be
  k ~=~ 
  \sum_{\nu,\alpha} \Big( 
  p_\alpha^\nu \oti \bar e_\nu^*(\bar q_\alpha^\nu) -
  e_\nu(p_\alpha^\nu) \oti \bar q_\alpha^\nu \Big)
  ~+~
  \sum_{\nu,\beta} \Big( 
  x_\beta^\nu \oti \bar e_\nu^*(\bar y_\beta^\nu) -
  e_\nu(x_\beta^\nu) \oti \bar y_\beta^\nu \Big) \ ,
\ee
where 
$p_\alpha^\nu \In P^+_s$, 
$\bar q_\alpha^\nu \In \bar P^{-\,*}_{p-s}$, and
$x_\beta^\nu \In P^{-}_{p-s}$, 
$\bar y_\beta^\nu \In \bar P^{+\,*}_{s}$, for $\nu = \pm$.
We have to show that
for any $k\in \Nc_s$ 
\be
  \Pi_s(k) = 0
  \quad \Rightarrow \quad
  k = 0 \ ,
\labl{eq:Pi-injective-statement}
where $\Pi_s$ is the projector defined just before (\ref{D1}). 
Since the image of $e_\nu$ lies in $N^\eps_s$, 
we have $\Pi_s \cir e_\nu = e_\nu$. Thus
\be
  \Pi_s(k) = 
  \sum_{\nu,\alpha} \Big( 
  \Pi_s(p_\alpha^\nu) \oti \bar e_\nu^*(\bar q_\alpha^\nu) -
  e_\nu(p_\alpha^\nu) \oti \bar q_\alpha^\nu \Big)
  +
  \sum_{\nu,\beta} \Big( 
  \Pi_s(x_\beta^\nu) \oti \bar e_\nu^*(\bar y_\beta^\nu) -
  e_\nu(x_\beta^\nu) \oti \bar y_\beta^\nu \Big) \ .
\labl{eq:Pi_s-to-general-el}
The summands in the sums over $\alpha$ and $\beta$ lie in different
direct summands of $\hat\Hc_s$, and so the equation $\Pi_s(k) = 0$
implies that both sums in \erf{eq:Pi_s-to-general-el} have to vanish
separately. Consider the first sum. We will prove below that
\be
  \sum_{\nu,\alpha} 
  \Pi_s(p_\alpha^\nu) \oti \bar e_\nu^*(\bar q_\alpha^\nu) -
  e_\nu(p_\alpha^\nu) \oti \bar q_\alpha^\nu = 0
  \quad \Rightarrow \quad
  \sum_{\nu,\alpha} 
  p_\alpha^\nu \oti \bar e_\nu^*(\bar q_\alpha^\nu) -
  e_\nu(p_\alpha^\nu) \oti \bar q_\alpha^\nu = 0 \ .
\labl{eq:Pi_s=0-first-half}
The corresponding statement for the second sum in
\erf{eq:Pi_s-to-general-el} can be seen analogously, and the two
statements together imply 
\erf{eq:Pi-injective-statement}, {\it i.e.}\ that $\Pi_s$ is injective
on $\Nc_s$.    
\medskip 

\noindent According to the decomposition \erf{eq:P-direct-sum} the
vectors $p_\alpha^\nu \in P^+_s$ can be written as
\be
p_\alpha^\nu
= v_\alpha^\nu + m_{\alpha,+}^{\nu} 
+ m_{\alpha,-}^{\nu} + u_\alpha^\nu \ ,
  \quad \text{where} \quad
  v_\alpha^\nu \In V^+_{s} \ , \quad
  m_{\alpha,\pm}^{\nu} \In S^+_{\pm,s} \ ,\quad
  u_\alpha^\nu \In U^+_{s} \ . 
\labl{eq:p-alpha-decomp}
Furthermore we have the induced decomposition of the dual spaces
$P_s^{\eps\,*} = V^{\eps\,*}_{s} \oplus S^{\eps\,*}_{+,s} \oplus 
S^{\eps\,*}_{-,s} \oplus U^{\eps\,*}_{s}$
and we will write $\bar q_\alpha^\nu$ as
\be
  \bar q_\alpha^\nu
  = \bar v_\alpha^\nu + \bar m_{\alpha,+}^{\nu} + 
    \bar m_{\alpha,-}^{\nu} + \bar u_\alpha^\nu \ ,
  \quad \text{where} \quad
  \bar v_\alpha^\nu \In \bar V^{-\,*}_{p-s} \ ,\quad 
  \bar m_{\alpha,\pm}^{\nu} \In \bar S^{-\,*}_{\pm,p-s} \ ,\quad
  \bar u_\alpha^\nu \In \bar U^{-\,*}_{p-s} \ . 
\labl{eq:q-alpha-decomp}
Here it is understood that, despite the similarity in notation,
$v_\alpha^\nu$ and $\bar v_\alpha^\nu$ are independent,
and similar for the other vectors in \erf{eq:p-alpha-decomp} 
and \erf{eq:q-alpha-decomp}.
Now for $e_\nu$ acting on $P^+_s$ we have
\be
\ker(e_\nu) = S^+_{\nu,s} \oplus U^+_s \ , \qquad
\ker(e_- \cir e_+) = S^+_{+,s} \oplus S^+_{-,s} \oplus U^+_s \ ,
\ee
and $e_\nu$ acts injectively on $V^+_s \oplus S^+_{-\nu,s}$, while
$n = e_- \cir e_+$ is injective on $V^+_s$. Dually, for $\bar e_\nu^*$
acting on $\bar P^{-\,*}_{p-s}$ we have
\be
\ker(\bar e_\nu^*) = \bar V^{-\,*}_{p-s} \oplus \bar
S^{-\,*}_{-\nu,p-s} \ , \qquad 
\ker(\bar e_+^* \cir \bar e_-^*) = 
\bar V^{-\,*}_{p-s} \oplus \bar S^{-\,*}_{+,p-s} 
\oplus \bar S^{-\,*}_{-,p-s} \ ,
\ee
$\bar e_\nu^*$ is injective on 
$\bar S^{-\,*}_{\nu,p-s} \oplus \bar U^{-\,*}_{p-s}$, and
$\bar n^* = \bar e_+^* \cir \bar e_-^*$ is injective on 
$\bar U^{-\,*}_{p-s}$.
Using these decompositions and kernels, we can write the condition
of the implication \erf{eq:Pi_s=0-first-half} as
\bea
  \sum_{\nu,\alpha} \Big(
  \big(m_{\alpha,+}^{\nu} + m_{\alpha,-}^{\nu} + u_\alpha^\nu)
     \oti \bar e_\nu^*\big(\bar m_{\alpha,\nu}^{\nu} + \bar u_\alpha^\nu\big) 
     \\[-.2em]\displaystyle \qquad \qquad -~
  e_\nu\big(v_\alpha^\nu + m_{\alpha,-\nu}^{\nu}\big) \oti 
     \big( \bar v_\alpha^\nu + \bar m_{\alpha,+}^{\nu} + 
    \bar m_{\alpha,-}^{\nu} + \bar u_\alpha^\nu \big) \Big)  ~=~ 0 \ .
\eear\labl{eq:lhs-is-zero-aux1}
When applying $e_{-\mu} \oti \bar n^*$ to this equation only the
second term of the sum survives and we obtain, for $\mu = \pm$,
\be
  \sum_{\alpha} e_{-\mu}\big(e_\mu( v_\alpha^\mu)\big) \oti
  \bar n^*(\bar u_\alpha^\mu) = 0
  \quad\Rightarrow\quad
  \sum_{\alpha} v_\alpha^\mu \oti \bar u_\alpha^\mu  = 0 ~,
\labl{eq:lhs-is-zero-aux2}
where in the implication we used that $n$ is injective on $V^+_s$
and $\bar n^*$ is injective on $\bar U^{-\,*}_{p-s}$. 
Applying $\id \oti \bar n^*$ to \erf{eq:lhs-is-zero-aux1} gives
\be
  \sum_{\nu,\alpha} 
  e_\nu(v_\alpha^\nu + m_{\alpha,-\nu}^{\nu}) \oti 
     \bar n^*( \bar u_\alpha^\nu) = 0
  \quad\Rightarrow\quad
  \sum_{\nu,\alpha} 
  e_\nu(m_{\alpha,-\nu}^{\nu}) \oti 
     \bar n^*( \bar u_\alpha^\nu) = 0 \ ,
\labl{eq:lhs-is-zero-aux3}
where the implication follows from the result \erf{eq:lhs-is-zero-aux2}.
Finally, applying $e_{-\mu} \oti \bar e_\mu^*$ to \erf{eq:lhs-is-zero-aux1}
results in
\be
  \sum_{\alpha} \Big(
  e_{-\mu}\big(m_{\alpha,\mu}^{-\mu})
     \oti \bar n^*\big(\bar u_\alpha^{-\mu}\big) -
  n\big(v_\alpha^\mu\big) \oti 
    \bar e_\mu^*\big(\bar m_{\alpha,\mu}^{\mu} +
    \bar u_\alpha^\mu \big) \Big)  ~=~ 0 \ .
\labl{eq:lhs-is-zero-aux4}
Summing this equation over $\mu = \pm$ and using \erf{eq:lhs-is-zero-aux3}
removes the first term, so that we are left with
\be
  \sum_{\alpha,\mu} 
  n\big(v_\alpha^\mu\big) \oti 
    \bar e_\mu^*\big(\bar m_{\alpha,\mu}^{\mu} +
    \bar u_\alpha^\mu \big) \Big) = 0
  \quad\Rightarrow\quad
  \sum_{\alpha,\mu} 
  v_\alpha^\mu \oti 
    \bar e_\mu^*\big(\bar m_{\alpha,\mu}^{\mu} +
    \bar u_\alpha^\mu \big)  = 0 \ ,
\labl{eq:lhs-is-zero-aux5}
where for the implication one uses that $n$ is injective on $V^+_s$.
Adding \erf{eq:lhs-is-zero-aux5} to \erf{eq:lhs-is-zero-aux1} gives
precisely the result of the implication \erf{eq:Pi_s=0-first-half},
thus completing the proof of \erf{eq:Pi_s=0-first-half}.


\end{document}